\documentclass[aps,pra,twocolumn,superscriptaddress]{revtex4-1}
\usepackage{bm}
\usepackage{graphicx}
\usepackage{amssymb,amsmath,amsbsy,amsgen,amsfonts}
\usepackage{dcolumn}
\usepackage{amsthm}
\usepackage{mathrsfs}
\usepackage{latexsym}
\usepackage{array}
\usepackage{amstext}
\usepackage{epsfig}
\usepackage{epstopdf} 

\usepackage{color}
\usepackage{units}
\usepackage{calrsfs}
\usepackage{verbatim}
\DeclareMathAlphabet\mathbfcal{OMS}{cmsy}{b}{n}

\newcommand{\be}{\begin{equation}}
\newcommand{\ee}{\end{equation}}
\newcommand{\ba}{\begin{array}}
\newcommand{\ea}{\end{array}}
\newcommand{\bqa}{\begin{eqnarray}}
\newcommand{\eqa}{\end{eqnarray}}

\begin{document}

\title{Optical torque on a two-level system near a strongly nonreciprocal medium}

\author{S. Ali Hassani Gangaraj}
\email{ali.gangaraj@gmail.com}
\address{School of Electrical and Computer Engineering, Cornell University, Ithaca, NY 14853, USA}

\author{ M\'ario G. Silveirinha}
\email{mario.silveirinha@co.it.pt}
\address{Instituto Superior T\'{e}cnico, University of Lisbon
	and Instituto de Telecomunica\c{c}\~{o}es, Torre Norte, Av. Rovisco
	Pais 1, Lisbon 1049-001, Portugal}

\author{George W. Hanson}
\email{george@uwm.edu}
\address{Department of Electrical Engineering, University of Wisconsin-Milwaukee, 3200 N. Cramer St., Milwaukee, Wisconsin 53211, USA}

\author{Mauro Antezza}
\email{mauro.antezza@umontpellier.fr}
\address{Laboratoire Charles Coulomb (L2C), UMR 5221 CNRS-Universit\'{e} de Montpellier, F-34095 Montpellier, France}
\address{Institut Universitaire de France, 1 rue Descartes, F-75231 Paris Cedex 05, France}

\author{Francesco Monticone}
\email{francesco.monticone@cornell.edu}
\address{School of Electrical and Computer Engineering, Cornell University, Ithaca, NY 14853, USA}

\date{\today}

\begin{abstract}

We investigate the quantum optical torque on an atom interacting with an inhomogeneous electromagnetic environment described by the most general linear constitutive relations.
The atom is modeled as a two-level system prepared in an arbitrary initial energy state. Using the Heisenberg equation of motion (HEM) and under the Markov approximation, we show that the optical torque has a resonant and non-resonant part, associated respectively with a spontaneous-emission process and Casimir-type interactions with the quantum vacuum, which can both be written explicitly in terms of the system Green function. Our formulation is valid for any inhomogeneous, dissipative, dispersive, nonreciprocal, and bianisotropic structure. We apply this general theory to a scenario in which the atom interacts with a material characterized by strong nonreciprocity and modal unidirectionality. In this case, the main decay channel of the atom energy is represented by the unidirectional surface waves launched at the nonreciprocal material-vacuum interface. To provide relevant physical insight into the role of these unidirectional surface waves in the emergence of non-trivial optical torque, we derive closed-form expressions for the induced torque under the quasi-static approximation. Finally, we investigate the equilibrium states of the atom polarization, along which the atom spontaneously tends to align due to the action of the torque. Our theoretical predictions may be experimentally tested with cold Rydberg atoms and superconducting qubits near a nonreciprocal material. We believe that our general theory may find broad application in the context of nano-mechanical and bio-mechanical systems. 

\end{abstract}

\maketitle


\section{Introduction}

Optically-induced mechanical torque, originating from the transfer of angular momentum between light and material bodies, is a topic of research with a long history, dating back to the early 20th century \cite{Beth,Francia,Allen}. The optical torque exerted on a trapped atom, molecule, or a micro-particle induces a rotation about a specific axis, and provides additional degrees of freedom to change the state of the considered object \cite{Rubinsztein-Dunlop,H. Chen}. This process leads to optical manipulations with a wide range of applications in physics, chemistry, biology, and medicine \cite{Ashkin,Padgett,Grier,Nori_1,Nori_2}. In a related context, Casimir effects due to zero-point energy fluctuations \cite{Lifshitz} are attracting increasing attention for their potential application in micro- and nano-electromechanical systems \cite{Capasso,Sirvent}. The Casimir effect was first discovered by H. Casimir in 1948 \cite{Casimir}, who predicted that two electrically-neutral metallic plates in vacuum experience an attractive force due to the zero-point fluctuations of the quantized electromagnetic field (i.e., due to the confinement of these fluctuations by the plates). In \cite{Weiss,Barash}, Casimir's theory was then generalized to the case of a system composed of two birefringent plates (i.e., a system with in-plane optical anisotropy) showing, for the first time, the emergence of a fluctuation-induced mechanical torque that makes the plates rotate toward a position with minimum zero-point energy. Further investigations and generalizations are discussed in \cite{Munday,Philbin,Mario_1,Mario_2,Buhmann_1,Buhmann_2,Buhmann_3,Buhmann_4}.


In a completely different context, it has been known for a few decades that optical systems with broken time-reversal symmetry may enable strong nonreciprocal wave-propagation effects. Most notably, suitably-designed nonreciprocal structures may support purely \emph{unidirectional} surface waves on the interface with a different medium \cite{1962}. A subclass of these systems that has been gaining increasing attention are ``photonic topological insulators'' with broken time-reversal symmetry -- the photonic analogue of the quantum-Hall insulators in condensed-matter physics \cite{Kane} -- in which unidirectional surface waves, emerging in the bulk-mode bandgap, are associated with a topological invariant number, a property that makes them intrinsically robust to continuous perturbations and immune to back-scattering at discontinuities \cite{Soljacic2014,Joannopoulos,Haldane}. We would like to stress, however, that the class of strongly nonreciprocal systems supporting unidirectional surface waves is broader than the class of nonreciprocal photonic topological insulators, as unidirectional surface modes may also exist outside the bulk-mode bandgap.  
For the purposes of our work, we therefore focus on unidirectional surface modes in general, rather than on modes with specific topological properties. In the context of optical forces/torques, it is indeed the possibility of having an asymmetric mode (unidirectional, in the extreme case) that may enable qualitatively different opto-mechanical effects compared to conventional optical structures. This has recently inspired new theoretical studies on the lateral recoil force exerted on an atom interacting with an interface that supports a unidirectional surface mode \cite{PRA_force,PRL_force}. In addition, the Casimir optical torque between two nonreciprocal topological surfaces was recently studied in \cite{Lindel}.
	

In contrast to any previous quantum-optical torque work, which focused on specific geometries in their ground energy state, in the present paper we offer a completely general theory of the quantum-optical torque exerted on a two-level system (e.g., an atom, molecule, quantum dot) interacting with a generic electromagnetic environment, composed of materials described by the most general linear constitutive relations. We consider both resonant and nonresonant contributions to the mechanical torque, associated, respectively, with the spontaneous emission of an initially-excited atom, and the fluctuation-induced Casimir torque. We then specialize our general theory to the case of an atom near a material with strong nonreciprocity, and provide relevant physical insight into unusual and counter-intuitive effects enabled by the unidirectional nature of the surface modes supported by this system.


The paper is organized as follows. In Sect. \ref{formulation}, we derive the optical quantum torque exerted on an atom in a generic dispersive bianisotropic electromagnetic environment. The derivation is based on the Heisenberg equation of motion, ignoring the effect of thermal fluctuations. The Markov approximation is used to derive a closed-form expression for the dynamics of the environment bosonic field and the atomic operators in the time domain. To simplify the problem, the atom is modeled as a two-level system. Using a quantized modal expansion, we show that, for an atom in an arbitrary initial state, the optical torque can be decomposed into a resonant and a non-resonant part. In Sect. \ref{Exact_sol}, the exact solution of the optical torque is written in terms of the system Green function, which allows further generalizing our theory to dissipative systems. In Sect. \ref{QS_sol}, we use our theory to study the case of a three-dimensional nonreciprocal material half-space, realized as a continuous gyrotropic material (a magnetized plasma). Then, under a quasi-static approximation, we characterize the plasmonic modes of the system and obtain an explicit expression for the optical torque. Finally, in Sect. \ref{Numerical_results}, we present an extensive numerical study that elucidates the symmetry requirements to obtain non-zero optical torque, and we explore the existence of equilibrium states along which the atom spontaneously tends to align due to the action of the optical torque.


\section{General theoretical formulation}
\label{formulation}

In this section, we investigate the quantum-optical torque on a two-level system, initially prepared in an arbitrary energy state, interacting with a generic inhomogeneous electromagnetic environment. The atom gets depopulated from its initial state through spontaneous emission. We rigorously formulate the optical torque that the atom experiences during the emission process in terms of the exact classical Green function of the system, and show that the expectation of the torque can be decomposed into a resonant term and a non-resonant (Casimir) term. 

\subsection{Modal Expansion of optical torque}

The quantum-optical torque operator is obtained from the classical definition of torque by promoting all quantities to operators \cite{torque},
\begin{equation}
	\hat{\boldsymbol{\tau}} = \hat{\textbf{p}}_g \times \hat{\textbf{F}},
\end{equation}
where the six-vector $ \hat{\textbf{F}} = [ \hat{\textbf{E}} ~~ \hat{\textbf{H}}    ]^{\mathrm{T}} $ contains the quantized electromagnetic fields of the system, and $ \hat{\textbf{p}}_g = [ \hat{\textbf{p}} ~~ \hat{\boldsymbol{0}} ]^{\mathrm{T}} $ is a generalized dipole moment operator, with zero magnetic dipole moment, such that $ \hat{\mathbf{p}} =\boldsymbol{\gamma}^* \hat{ \sigma}_+ + \boldsymbol{\gamma} \hat{ \sigma}_- $, where $\boldsymbol{\gamma} $ is the dipole transition matrix element and $\hat{\sigma}_{\pm }$ the atom raising and lowering operators. As usually done, the quantized electromagnetic fields can be written in the form of positive and negative frequency components $ \hat{\textbf{F}} = \hat{\textbf{F}}_- + \hat{\textbf{F}}_+ $, with $ \hat{\textbf{F}}_+ = \hat{\textbf{F}}_-^{\dagger} $ due to the reality (Hermiticity) of the operator. We then expand the negative-frequency quantized field as
\begin{equation}\label{F-}
	\hat{\textbf{F}}_- (\textbf{r},t) = \sum_{\omega_{n\textbf{k}}>0} \sqrt{ \frac{\hbar \omega_{n\textbf{k}}  }{2}  } \textbf{F}_{n\textbf{k}} (\textbf{r}) \hat{a}_{n\textbf{k}}(t),
\end{equation}
where $ \textbf{F}_{n\textbf{k}} (\textbf{r})  $ represents a cavity mode normalized as \cite{Mario_N_SE}
\begin{equation}
	\frac{1}{2}\int\limits_{{}}{{d^{3}}\mathbf{r}\,\mathbf{F}_{n\mathbf{k}%
		}^{\ast }\cdot \frac{{\partial \left( {\omega \mathbf{M}}\right) }}{{%
			\partial \omega }}\cdot \,}\mathbf{F}_{n\mathbf{k}}^{{}}=1,  \label{norm}
\end{equation}%
where $ \omega_{n\textbf{k}} $ is the mode oscillation frequency, and  
\begin{equation}
\mathbf{M}=\left( {%
	\begin{array}{ccccccccccccccccccc}
	{\boldsymbol{\varepsilon }\left({\mathbf{r},\omega }\right) } & \boldsymbol{\xi}\left({\mathbf{r},\omega }\right)/c \\
	\boldsymbol{\zeta}\left({\mathbf{r},\omega }\right)/c & { {\mu} }\left({\mathbf{r},\omega }\right)
	\end{array}%
}\right) 
\end{equation}%
contains the material information, which relates the classical $\bf D$ and $\bf B$ fields with the classical $\bf E$ and $\bf H$ fields. Eq. (\ref{norm}) indicates that the stored energy of the modes is normalized to unity. In the modal expansion (\ref{F-}), $ \hat{a}_{n\textbf{k}} $ is the (annihilation) bosonic field operator ($\hat{a}_{n%
	\mathbf{k}}^{\dagger }$ is the corresponding creation operator), which represents the reservoir field and satisfies $  [ \hat{a}_{n\textbf{k}}, \hat{a}_{n\textbf{k}'}^{\dagger} ] = \delta_{\textbf{k} , \textbf{k}'} $. 

The total Hamiltonian of the system in the dipole approximation is
\begin{align}
\hat{ \textbf{H}} =& \hbar \omega _{0}\hat{\sigma}_{+}\hat{\sigma}%
_{-}+\sum_{\omega _{n\mathbf{k}}>0}\frac{\hbar \omega _{n\mathbf{k}}}{2}%
\left( \hat{a}_{n\mathbf{k}}\hat{a}_{n\mathbf{k}}^{\dagger }+\hat{a}_{n%
	\mathbf{k}}^{\dagger }\hat{a}_{n\mathbf{k}}\right) \notag \\ &
  -\hat{\mathbf{p}}\cdot \hat{\mathbf{E}}(\mathbf{{r}_{0}}) 
\end{align}
where the right side can be decomposed into the atom Hamiltonian (first term), the reservoir Hamiltonian (second term) and the interaction Hamiltonian (third term, where $ \textbf{r}_0 $ is the atom position). Using the Heisenberg equation of motion $\partial _{t}\hat{a}_{n\mathbf{k}}=i\hbar ^{-1}\left[ \hat{ \textbf{H}},\hat{a}_{n\mathbf{k}}\right] $ for a single atom interacting with reservoir electric field and applying the Markov approximation (i.e., the evolution of the quantum system is ``local'' in time; at each time instant $t$, the system's memory of earlier times $ t<t' $ is negligible), it can be shown that \cite{Eberly}
\begin{align}\label{a-hat}
	\hat{a}_{n\mathbf{k}}(t)\approx &  \hat{a}_{n\mathbf{k}}e^{-i\omega _{n \mathbf{k}}t}  \notag \\ &   +\sqrt{\frac{\omega _{n\mathbf{k}}}{2\hbar }}\tilde{\boldsymbol{\gamma }} \cdot {\mathbf{F}}_{n\mathbf{k}}^{\ast
	}(\mathbf{r}_0)\hat{\sigma}_{-}(t)\frac{1}{\omega
		_{n\mathbf{k}}-\omega _{0}} \notag \\ &  +\sqrt{\frac{\omega _{n\mathbf{k}}}{2\hbar }}\tilde{\boldsymbol{\gamma }}^{\ast }\cdot {\mathbf{F}}_{n\mathbf{k}}^{\ast }(\mathbf{r}_0)\hat{\sigma}_{+}(t)\frac{1}{ \omega _{n\mathbf{k}}+\omega _{0}}  
\end{align}
where $ \tilde{\boldsymbol{\gamma}} = [ \boldsymbol{\gamma}~~ 0   ]^{\mathrm{T}} $. The first term $  \hat{a}_{n\mathbf{k}}e^{-i\omega _{n \mathbf{k}}t}  $ is the free-field solution without an emitter. Using normal ordering, the expectation of the torque can be written as $\hat{ \mathbf{\tau}} = 2 \mathrm{Re} \left <  \hat{ \textbf{p}}_g \times  \hat{\textbf{F}}_-  \right> $ (note that, by writing the torque in terms of $ \hat{\textbf{F}}_- $, the free-field operator does not contribute to the final expression of the torque, as $\hat{\textbf{F}}_-$ acts on the vacuum state, and $\hat{a}_{n\mathbf{k}} {\left | 0 \right >} =0$). Then, using Eqs. (\ref{F-}) and (\ref{a-hat}) and considering 
\begin{align}
	& \left< \hat{ \sigma}_+(t) \hat{ \sigma}_-(t)    \right> = \rho_{ee}(t), ~ \left< \hat{ \sigma}_-(t) \hat{ \sigma}_+(t)    \right> = 1 - \rho_{ee}(t) \notag \\ & \left< \hat{ \sigma}_+(t) \hat{ \sigma}_+(t)    \right> = \left< \hat{ \sigma}_-(t) \hat{ \sigma}_- (t)   \right> = 0,
\end{align}
where $ \rho_{ee}(t) $ is the probability of the atom to be found in its excited state, one can finally obtain, 
\begin{align}\label{tau_hat}
& \hat{ \boldsymbol{\tau}} = \rho_{ee}(t) \hat{ \boldsymbol{\tau}}_r + (1 - \rho_{ee}(t)) \hat{ \boldsymbol{\tau}}_n,
\end{align}
where $ \hat{ \mathbf{\tau}}_r $ and $ \hat{ \mathbf{\tau}}_n $ are the resonant and non-resonant parts of the optical torque, respectively, which can be written in terms of the modal expansion as
\begin{align}\label{tau}
	& \hat{ \boldsymbol{\tau}}_r  = \mathrm{Re} \left \{ \sum_{n \mathbf{k} > 0 }     \tilde{\boldsymbol{\gamma}}^* \times \left[\left( \mathbf{F}_{ n \mathbf{k} }(\mathbf{r}_0) \otimes  \mathbf{F}_{n\mathbf{k}}^*(\mathbf{r}_0)  \right) \cdot \tilde{\boldsymbol{\gamma}}\right]  \frac{ \omega_{n \mathbf{k}} }{ \omega_{n\mathbf{k}} - \omega_0 } \right \}  \notag \\&
	\hat{ \boldsymbol{\tau}}_n = \mathrm{Re}  \left \{\sum_{n \mathbf{k} > 0 }     \tilde{\boldsymbol{\gamma}} \times\left[ \left( \mathbf{F}_{ n \mathbf{k} }(\mathbf{r}_0) \otimes \mathbf{F}^*_{n\mathbf{k}}(\mathbf{r}_0)  \right) \cdot \tilde{\boldsymbol{\gamma}}^*\right]  \frac{  \omega_{n \mathbf{k}} }{ \omega_{n\mathbf{k}} + \omega_0 } \right \}.
\end{align} 

In the non-pumped scenario, in which the atom is assumed to be initially prepared in an excited state, and there is no external excitation during the spontaneous emission process, the excited state population decays exponentially as $ \rho_{ee}(t) = e^{-\Gamma_{11}t} $, where $ \Gamma_{11} $ is the spontaneous emission rate. \\


\subsection{Exact solution in terms of the Green function}
\label{Exact_sol}

In this section, we express the modal expansion of the torque in Eqs. (\ref{tau}) in terms of the Green function of the system. The classical Green function for the wave equation satisfies $ \textbf{N} \cdot \textbf{G} = \omega \textbf{M} \cdot \textbf{G} + i \textbf{I}\delta(\textbf{r} - \textbf{r}_0) $, where $\textbf{N} =\left( \begin{array}{cc}
 \textbf{0} & i \boldsymbol{\nabla} \times \textbf{I}_{3 \times 3} \\
 -i \boldsymbol{\nabla} \times \textbf{I}_{3 \times 3} & \textbf{0}
\end{array}\right)$ contains the spatial derivatives, and  $ \textbf{G} =\left( \begin{array}{cc}
\textbf{G}_{\mathrm{EE}} & \textbf{G}_{\mathrm{EH}} \\
\textbf{G}_{\mathrm{HE}} & \textbf{G}_{\mathrm{HH}}
\end{array}\right) $ is a $ 6 \times 6 $ dyad. It can be shown that, in the limit of low losses, the Green function of the system can be written in term of the modal expansion as follows \cite{Mario_N_SE}
\begin{equation}
\mathbf{G} = \mathbf{G}^+ + \mathbf{G}^- + \frac{1}{i \omega} \mathbf{M}^{-1}_{\infty}\delta(\mathbf{r} - \mathbf{r}_0),
\end{equation}
where 
\begin{align}\label{Gmp}
{\left( {-i\omega }\right) \mathbf{G}}^{+}& =\sum\limits_{{%
		\omega _{n\mathbf{k}}}>0}{\frac{{{\omega _{n\mathbf{k}}}}}{{2}}\frac{1}{%
		\omega _{n\boldsymbol{\mathrm{k}}}-\omega }}\boldsymbol{%
	\mathrm{F}}_{n\boldsymbol{\mathrm{k}}}(\boldsymbol{\mathrm{r}})\otimes
\boldsymbol{\mathrm{F}}_{n\boldsymbol{\mathrm{k}}}^{\ast }(\boldsymbol{%
	\mathrm{r}}_{0})  \nonumber \\
{\left( {-i\omega}\right) \mathbf{G}}^{-}& =\sum\limits_{{%
		\omega _{n\mathbf{k}}}>0}{\frac{{{\omega _{n\mathbf{k}}}}}{{2}}\frac{1}{%
		\omega _{n\boldsymbol{\mathrm{k}}}+\omega }}\boldsymbol{%
	\mathrm{F}}_{n\boldsymbol{\mathrm{k}}}^{\ast }(\boldsymbol{\mathrm{r}}%
)\otimes \boldsymbol{\mathrm{F}}_{n\boldsymbol{\mathrm{k}}}(\boldsymbol{%
	\mathrm{r}}_{0})
\end{align}%
are the positive and negative frequency parts of the Green function, respectively, and $ \mathbf{M}_{\infty} $ is the material response at infinite frequency. The $\delta$-function term is associated with the self-field, which we ignore \cite{torque}. Then, by comparing Eq. (\ref{Gmp}) and Eqs. (\ref{tau}), we can write (\ref{tau_hat}) as
\begin{align}
\hat{ \mathbf{\tau}}  = & 2 \rho_{ee}(t)  \mathrm{Re} \left \{  \tilde{\boldsymbol{\gamma}}^* \times \left( -i\omega \mathbf{G}^+(\mathbf{r}_0, \mathbf{r}_0, \omega_0)  \right)  \cdot \tilde{\boldsymbol{\gamma}}  \right \}  \notag \\ & + 2 (1 - \rho_{ee}(t))  \mathrm{Re} \left \{  \left[\tilde{\boldsymbol{\gamma}}^* \times \left( -i\omega \mathbf{G}^-(\mathbf{r}_0, \mathbf{r}_0, \omega_0)  \right)  \cdot \tilde{\boldsymbol{\gamma}}\right]^*  \right \}.
\end{align} 

The conjugation of the second term in the above equation can be safely dropped, since we are taking the real part. We then replace $ \mathbf{G}^+ $ by $ \mathbf{G} - \mathbf{G}^- $, and by noting that $\mathbf{G}^- $, defined in (\ref{Gmp}), is a complex analytic function for ${\mathop{\rm Re}\nolimits}\left\{ \omega  \right\} > 0$, we can invoke Cauchy's theorem and represent the second term of this equation as an integral over the imaginary frequency axis,
\begin{align}\label{tau_int}
\hat{ \mathbf{\tau}}  = & 2 \rho_{ee}(t)  \mathrm{Re} \left \{  \tilde{\boldsymbol{\gamma}}^* \times \left( -i\omega \mathbf{G}(\mathbf{r}_0, \mathbf{r}_0, \omega_0)  \right)  \cdot \tilde{\boldsymbol{\gamma}}  \right \} \notag \\ &  + 2 (1 - 2\rho_{ee}(t)) \,{\mathop{\rm Re}\nolimits} \left\{
{\frac{1}{{2\pi }}\int\limits_{ - \infty }^\infty  {d\xi~ }
	{{\bf{\tilde \gamma }}^*} \times \frac{{{{\left( { - i\omega {{ {\bf{G}} }^ - }} \right)}_{\omega  = i\xi
	}}}}{{{\omega _0} - i\xi }} \cdot {\bf{\tilde \gamma }}} \right\}.
\end{align} 
Since $  {\int\limits_{ - \infty }^\infty  {d\xi~ }
		{{\bf{\tilde \gamma }}} \times \frac{{{{\left( { - i\omega {{ {\bf{G}} }^ - }} \right)}_{\omega  = i\xi
		}}}}{{{\omega _0} + i\xi }} \cdot {\bf{\tilde \gamma }}^*}=0 $ (no poles are enclosed by the Cauchy's integration contour with an infinite semicircle for ${\mathop{\rm Re}\nolimits}\left\{ \omega  \right\} > 0$), we can add this term to the integrand above without changing the result, 
	\begin{widetext}
		\begin{align}\label{compare}
		\hat{ \mathbf{\tau}}  = & 2 \rho_{ee}(t)  \mathrm{Re} \left \{  \tilde{\boldsymbol{\gamma}}^* \times \left( -i\omega \mathbf{G}(\mathbf{r}_0, \mathbf{r}_0, \omega_0)  \right)  \cdot \tilde{\boldsymbol{\gamma}}  \right \} \notag \\ &  + 2 (1 - 2\rho_{ee}(t)) \,{\mathop{\rm Re}\nolimits} \left\{
		\frac{1}{{2\pi }}\int\limits_{ - \infty }^\infty  {d\xi }\left[
		{{\bf{\tilde \gamma }}^*} \times \frac{{{{\left( { - i\omega {{ {\bf{G}} }^ - }} \right)}_{\omega  = i\xi
		}}}}{{{\omega _0} - i\xi }} \cdot {\bf{\tilde \gamma }}  + {{\bf{\tilde \gamma }}} \times \frac{{{{\left( { - i\omega {{ {\bf{G}} }^ - }} \right)}_{\omega  = i\xi
		}}}}{{{\omega _0} + i\xi }} \cdot {\bf{\tilde \gamma^* }}\right] \right\}. 
		\end{align} 
	\end{widetext}
Moreover, since we take the real part of the non-resonant term, we can conjugate the integrand, and noting that ${\left( { - i\omega {{\overline {\bf{G}} }^ -}\left( {{{\bf{r}}_0},{{\bf{r}}_0},\omega } \right)} \right)_{\omega = i\xi }} = {\left[ {{{\left( { - i\omega {{\overline {\bf{G}} }^ +}\left( {{{\bf{r}}_0},{{\bf{r}}_0},\omega } \right)} \right)}_{\omega  = i\xi }}} \right]^*}$, we can then replace ${{\bf{G}}^ - }$ by ${{\bf{G}}^ + }$. This implies that we can also replace ${{\bf{G}}^ - }$ in Eq. (\ref{compare}) by one-half of the full Green function, ${{\bf{G}}}/2$, without changing the result. We therefore obtain
\begin{widetext}
\begin{align}\label{tau_exact}
\hat{ \mathbf{\tau}}  = & 2 \rho_{ee}(t)  \mathrm{Re} \left \{  \tilde{\boldsymbol{\gamma}}^* \times \left( -i\omega \mathbf{G}(\mathbf{r}_0, \mathbf{r}_0, \omega_0)  \right)  \cdot \tilde{\boldsymbol{\gamma}}  \right \} \notag \\&
 + 2 (1 - 2\rho_{ee}(t)) \,{\mathop{\rm Re}\nolimits} \left\{
\frac{1}{{4\pi }}\int\limits_{ - \infty }^\infty  {d\xi }
\left[{{\bf{\tilde \gamma }}^*} \times \frac{{{{\left( { - i\omega {{ {\bf{G}} } }} \right)}_{\omega  = i\xi
}}}}{{{\omega _0} - i\xi }} \cdot {\bf{\tilde \gamma }}  + {{\bf{\tilde \gamma }}} \times \frac{{{{\left( { - i\omega {{ {\bf{G}} } }} \right)}_{\omega  = i\xi
}}}}{{{\omega _0} + i\xi }} \cdot {\bf{\tilde \gamma^* }}\right] \right\},
\end{align} 
\end{widetext}
which gives the expression of the quantum-optical torque in terms of the classical Green function of the system, for an atom interacting with a generic inhomogeneous bianisotropic electromagnetic reservoir. While in the derivation above we have initially made the assumption of vanishing losses in order to write the fields as a sum of orthonormal cavity modes in Eq. (\ref{F-}), the final expression in Eq. (\ref{tau_exact}) is given in terms of the full system's Green function, not a modal expansion. Thanks to this formulation, the first term of Eq. (\ref{tau_exact}) is valid in all cases, including for lossy systems, whereas the second term only requires some minor modifications in the case of a lossy environment, as detailed in Appendix \ref{zero_point_lossy}.

More broadly, our general theory applies to any lossy and/or active media, and hence it can directly be extended to non-Hermitian nonreciprocal/topological systems, which are currently the subject of several studies (e.g., \cite{Zhen,NHT}). The analysis of non-Hermitian topological media as reservoirs for quantum emitters has recently been studied in \cite{PTI_C(t)}, and will be considered further in future works.


\begin{figure}[!htbp]
	\begin{center}
		\noindent \includegraphics[width=2.1in]{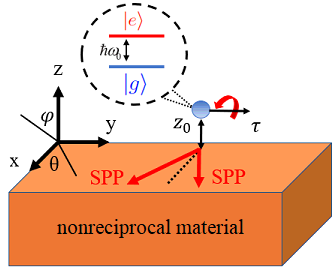}
	\end{center}
	\caption{A two-level system (e.g., an atom) near the surface of a nonreciprocal (gyrotropic) material, experiencing an optical torque. The black arrow indicates the direction of the torque, and the curved red arrow represents the direction of the induced rotation in the $ xoz $-plane. The radiation emitted by an initially-excited atom may launch \emph{unidirectional} surface waves (e.g., surface plasmon polaritons, SPPs) on the surface of the nonreciprocal material, resulting in a resonant optical torque. An additional non-resonant contribution to the optical torque originates from quantum-vacuum fluctuations (Casimir torque), which exists even for an atom in its ground state.}
	\label{geom}
\end{figure}

\section{Quasi-static analysis in a three-dimensional nonreciprocal material environment}
\label{QS_sol}

To provide a closed-form quasi-static evaluation of Eq. (\ref{tau_exact}), we need to specialize our discussion to a system of interest, and consider the relevant modes supported by the structure that contribute to the torque exerted on the atom. We consider here a scenario in which the electromagnetic environment is a three-dimensional nonreciprocal material half-space. As shown in Fig. \ref{geom}, the considered system is stratified in the $z$ direction, with a generic continuum nonreciprocal material filling the half-space $ z < 0 $, and vacuum in the upper half-space $ z>0 $. An atom modeled as a two-level system with Hamiltonian $ \hat{ \textbf{H}}_a = \sum_{j=e,g} \mathrm{E}_j \left | j \right > \left < j \right | $, is located at $ z = z_0 $ above the interface, where $ \mathrm{E}_j $ is the energy associated with the excited and ground states of the atom. As a specific example of a continuum nonreciprocal medium, we consider a gyrotropic material with tensorial permittivity $ \boldsymbol{\epsilon} = \epsilon_0 \left(  \epsilon_t \textbf{I}_t + \epsilon_a \hat{ \textbf{y}} \hat{ \textbf{y}} + i \epsilon_g \hat{ \textbf{y}} \times \textbf{I}  \right) $ and $ \boldsymbol{\mu} = \mu_0 \textbf{I} $, where $ \textbf{I}_t = \textbf{I} - \hat{ \textbf{y}} \hat{ \textbf{y}} $, and $ \epsilon_g $ is a gyrotropy parameter that models the nonreciprocal response of the material (the permittivity tensor is non-symmetric if $ \epsilon_g \neq 0 $).

When the atom is located electrically close to the gyrotropic material, $ z_0 \ll \lambda_0 $, where $ \lambda_0 $ is the free-space radiation wavelength, the dynamics of the system is expected to be governed by the surface waves excited on the nonreciprocal material-air interface, e.g., surface plasmon polaritons (SPPs) if the gyrotropic material is a magnetized plasma. This allows us to replace the modal expansion in Eqs. (\ref{tau}) with its quasi-static solution $\boldsymbol{\mathrm{F}}_{n\mathbf{k}}\approx \left[ \boldsymbol{\mathrm{E}}_{n\mathbf{k}}~~\boldsymbol{0}\right] ^{\mathrm{T}}\approx \left[ -\nabla \phi _{\mathbf{k}}~~ \boldsymbol{0}\right] ^{\mathrm{T}}$. The magnetic field is assumed negligible, and the electric field is written in terms of an electric potential $\phi _{\mathbf{k}}$ that satisfies Laplace's equation, $\nabla \cdot (\boldsymbol{\varepsilon }\cdot \nabla \phi _{\mathbf{k}})=0$. The solutions of this quasi-static equation are of the form
	
\begin{equation}
\phi _{\mathbf{k}}=\frac{{A}_{\mathbf{{k}_{\parallel }}}}{\sqrt{S}}%
e^{i\mathbf{{k}_{\parallel }}\cdot \boldsymbol{\mathrm{r}}}%
\begin{cases}
e^{-{k}_{\parallel }z}, & z>0 \\
e^{+\tilde{{k}}_{\parallel }z}, & z<0%
\end{cases}
\label{elecpot2}
\end{equation}%
where ${{\bf{k}}_{\parallel}} = {k_x}{\bf{\hat x}} + {k_y}{\bf{\hat y}}$ is the wavevector of the SPPs, $ \tilde{k}_{\parallel} = \sqrt{k_x^2 + (\epsilon_a/\epsilon_t)k_y^2 } $, $S$ is the area of the slab, and 
\begin{equation}
|{A}_{\boldsymbol{{\rm k}}_{\parallel }}|^{2}=\frac{2}{\varepsilon
	_{0}}\left[ {{k}}_{\parallel }+\frac{\Lambda (\omega _{\theta
	},\omega _{c},\omega _{p})}{2\tilde{{{k}}}_{\parallel }}\right]
^{-1}
\end{equation}%
with $\Lambda (\omega ,\omega _{c},\omega _{p})=\partial _{\omega
}\left(
\varepsilon _{t}\omega \right) \left( \tilde{{{k}}}_{\parallel }^{2}+%
{k}_{x}^{2}\right) +\partial _{\omega }\left( \varepsilon _{a}\omega
\right) {k}_{y}^{2}+\partial _{\omega }\left( \varepsilon _{g}\omega
\right) 2 {k}_{x}\tilde{{{k}}}_{\parallel }$. By applying the boundary condition at the interface (i.e., $\mathbf{\hat{z}}\cdot \mathbf{\varepsilon }\cdot \nabla {\phi _{\mathbf{k}}}$ continuous at $z=0$), we find the SPP dispersion equation
\begin{equation}\label{SPPresonance}
-{k}_{\parallel }={k}_{x}\varepsilon _{g}(\omega )+\tilde{%
	{k}}_{\parallel }\varepsilon _{t}(\omega ).
\end{equation}

For a lossy magnetized plasma with bias magnetic field along the $ +y $-axis, the frequency-dispersive permittivity elements are given by \cite{Bittencourt}

\begin{align}
\label{bp} &
{\varepsilon _t} = 1 - \frac{{\omega _p^2\left( {1 + i\Gamma /\omega } \right)}}{{{{\left( {\omega  + i\Gamma } \right)}^2} - \omega _c^2}} \nonumber \\
& {\varepsilon _a} = 1 - \frac{{\omega _p^2}}{{\omega \left( {\omega
			+ i\Gamma } \right)}}, \,\,\,\, {\varepsilon _g} = \frac{1}{\omega
}\frac{{\omega _c^{}\omega _p^2}}{{\omega _c^2 - {{\left( {\omega  +
					i\Gamma } \right)}^2}}},
\end{align}
where $\omega _{p}$ is the plasma frequency, $\Gamma$ the collision rate associated with damping, $\omega _{c}=-q|B_{0}|/m$ the cyclotron frequency, $q=-e$ the electron charge, $m$ the effective electron mass, and $B_{0}$ the static bias. As an example of a material platform, $n$-doped narrow-gap semiconductors such as $n$-type InSb under moderate magnetic bias act as magnetized electron gases, with permittivity elements consistent with (\ref{bp}) in the low THz range \cite{Palik, GarciaVidal}. Considering the dispersive material model of biased plasma in the limit of zero damping, the solution of Eq. (\ref{SPPresonance}) yields a single branch of modes $\omega_{\boldsymbol{\mathrm{k}}}$, which depends only on the angle
${\theta}$ of the wavevector with respect to the $+x$-axis,
\begin{equation}  \label{w_theta}
\omega_{\boldsymbol{\mathrm{k}}}=\omega_{\theta} = \frac{%
	\omega_c}{2} ~ \mathrm{cos} ({\theta}) + \sqrt{ \frac{\omega_p^2%
	}{2} + \frac{\omega_c^2}{4} (1 + \mathrm{sin}^2({\theta})) }.
\end{equation}
\begin{figure}[!htbp]
	\begin{center}
		\noindent \includegraphics[width=2.7in]{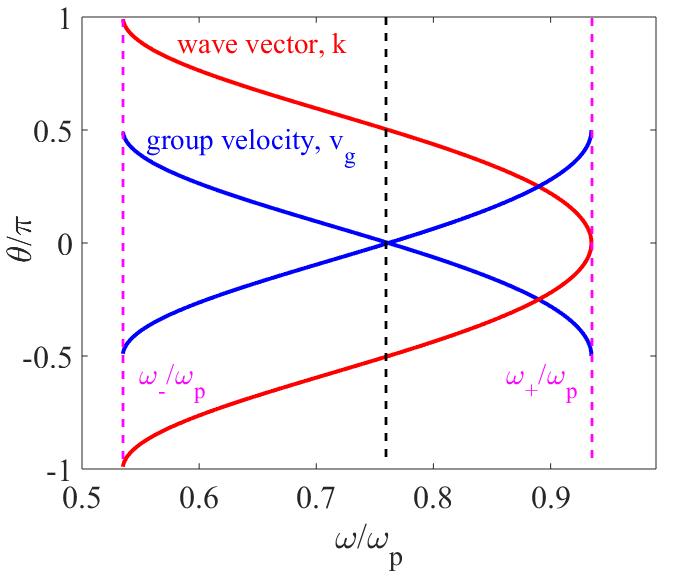}
	\end{center}
	\caption{Angular direction of the dominant wavevector (red) and group velocity (direction of energy flow) (blue) of the two SPP beams launched on the surface of the nonreciprocal material in Fig. \ref{geom} (biased plasma), as a function of frequency for $ \omega_c = 0.4 \omega_p $. The dashed pink lines indicates $ \omega_- $ and $ \omega_+ $, defined in the text. In this frequency range, the equifrequency contours of the dispersion function are hyperbolic, with the dominant wavevector of the beams determined by the hyperbola asymptotes. In comparison, for a typical reciprocal hyperbolic material, the figure would have four branches for the wavevector and group velocity, instead of two, due to the symmetry of the equifrequency contour with respect to the origin of $\boldsymbol{\mathrm{k}}$-space.}
	\label{w_t}
\end{figure}

For $\omega_c>0$, one has $\omega_- < \omega_{\boldsymbol{\mathrm{k}}} < \omega_+ $%
, with
\begin{align}  \label{w+_w-}
& \omega_+ \equiv \omega_{{k}_x >0, {k}_y = 0} = \frac{1}{2}
\left( \omega_c + \sqrt{2\omega_p^2 + \omega_c^2} \right),  \nonumber \\
& \omega_- \equiv \omega_{{k}_x <0, {k}_y = 0} = \frac{1}{2} \left(
-\omega_c + \sqrt{2\omega_p^2 + \omega_c^2} \right) .
\end{align}

In this frequency range, the equifrequency contours of the dispersion function of the SPP are hyperbolic curves. The hyperbolic dispersion results in two beams at $ \pm \theta $ with respect to +$x$-axis \cite{PRL_force,PRA_force}. It is important to note that the quasi-static analysis above gives the SPP solutions for large wavenumbers (i.e., short wavelengths), which contain the dominant part of the atom-environment interaction. Fig. \ref{w_t} shows the angle of the dominant wavevector (red lines) and group velocity vector (blue line) of the launched hyperbolic SPP beams with respect to the $+x$-axis. Note that the group velocity direction (power flow direction) is approximately rotated by 90 degrees with respect to the wavevector direction, consistent with an hyperbolic equifrequency contour for wavevectors of large magnitude. We also note that as frequency varies $ \omega_-  \rightarrow \omega_+ $, the wavevector sweeps the entire $xoy$ plane, whereas the direction of power flow always remains in the half-plane that includes the positive $x$-axis. This asymmetry in the excitation of SPPs (no energy launched toward the negative $x$-semiplane) is a consequence of breaking reciprocity by applying a static magnetic bias, which enables the emergence of unidirectional surface modes. We would like to stress that such unidirectionality is crucial to have non-trivial optical torque. The torque exerted on an atom is associated with a recoil force; due to conservation of momentum, the direction of this force is opposite to the direction of momentum release (wavevector). Therefore, in reciprocal systems in which SPP propagation is symmetric in space, the symmetric release of momentum does not lead to any net force/torque on the emitter. Instead, nonreciprocity allows breaking these symmetries, hence leading to non-trivial optical torque \cite{Note}.

In the following subsections, we use the quasi-static solution of the eigenmodes in Eq. (\ref{elecpot2}) and the corresponding eigenfrequencies in Eq. (\ref{w_theta}) to derive a closed-form expression for the resonant and non-resonant parts of the optical torque in Eqs. (\ref{tau}).


\subsection{Resonant term}

The resonant component of the torque is given by the first of Eqs. (\ref{tau}). Rewriting this for a single mode and considering $ \mathbf{ F}_{n\mathbf{ k}} \approx \mathbf{ F}_{\mathbf{ k}} = [\mathbf{ E}_{\mathbf{ k}} ~ \mathbf{ 0}]^{\mathrm{T}} = [- \nabla { \phi}_{\mathbf{ k}} ~ \mathbf{ 0}]^{\mathrm{T}} $ gives
	\begin{align}
	&
	\mathbf{F}_{\mathbf{k}}(\mathbf{r}) = \left[  -( i\mathbf{k}_{\parallel} - \mathrm{k}_{\parallel} \hat{z} ) \frac{ A_{\mathbf{k}_{\parallel}  }}{ \sqrt{S} } e^{ i \mathbf{k}_{\parallel} \cdot \mathbf{r} - \mathrm{k}_{\parallel} z } ~~~ \mathbf{0}\right]^{\mathrm{T}} .
	\end{align}
Considering the translational invariance of the system in the $xoy$-plane, we can use use polar coordinates $ \mathbf{k}_{\parallel} = \mathrm{k}_{\parallel} ( \mathrm{cos}(\theta),~ \mathrm{sin}(\theta), ~ 0 ) $ and, by replacing $ \omega_{\mathbf{k}} $ with $ \omega_{\theta} $, we can transform the summation over the discrete modes in (\ref{tau}) into an integral, $\frac{1}{S} \sum\limits_{{\omega _{\mathbf{k}}}>0}\rightarrow \frac{1}{{{{\left( {2\pi } \right) }^{2}}}} \int_{\theta = 0}^{ \theta = 2\pi } d\theta \int_{\mathrm{k}_{\parallel} = 0}^{\mathrm{k}_{\parallel} = +\infty} \mathrm{k}_{\parallel} d \mathrm{k}_{\parallel} $, obtaining
\begin{widetext}
\begin{align}\label{tau-r2}
	\hat{ \boldsymbol{\tau}}_r = \mathrm{Re} \left \{  \frac{1}{{{{\left( {2\pi } \right) }^{2}}}} \int_{\theta = 0}^{ \theta = 2\pi } d\theta \int_{\mathrm{k}_{\parallel} = 0}^{\mathrm{k}_{\parallel} = +\infty} d \mathrm{k}_{\parallel} \mathrm{k}_{\parallel}  |A_{\mathbf{k}_{\parallel} }|^2  e^{-2 k_{\parallel} z_0}  \omega_{ \theta }  \boldsymbol{\gamma}^* \times \left[ ( i\mathbf{k}_{\parallel} - \mathrm{k}_{\parallel} \hat{z} )( -i\mathbf{k}_{\parallel} - \mathrm{k}_{\parallel} \hat{z} ) \cdot \boldsymbol{\gamma}   \right]   \frac{1} { \omega_{\theta} - \omega_0}  \right \}.
\end{align}
\end{widetext}
The integral in (\ref{tau-r2}) can be re-written as 
\begin{align}\label{tau-2}
	\hat{ \boldsymbol{\tau}}_r =  \mathrm{Re} & \left \{   \frac{1}{ \epsilon_0 {{{\left( {2\pi } \right) }^{2}}}} \int_{\theta = 0}^{ \theta = 2\pi } d\theta \int_{\mathrm{k}_{\parallel} = 0}^{\mathrm{k}_{\parallel} = +\infty}   \mathrm{k}_{\parallel}^2  e^{-2 k_{\parallel} z_0} d \mathrm{k}_{\parallel}  \right. \nonumber \\&
\left.    a_{\theta} \omega_{ \theta }  \left[\boldsymbol{\gamma}^* \times \mathbf{M}_{\theta}\right]  \frac{1} { \omega_{\theta} - \omega_0}  \right \},
\end{align}
where ${a_\theta } \equiv {\left| {{A_{{{\bf{k}}_{\parallel}}}}} \right|^2}{\varepsilon _0}k_{\parallel}^{}$ and 
\begin{align}
	\mathbf{M}_{\theta} & = \frac{1}{k_{\parallel}^2} ( i\mathbf{k}_{\parallel} - \mathrm{k}_{\parallel} \hat{z} )( -i\mathbf{k}_{\parallel} - \mathrm{k}_{\parallel} \hat{z} ) \cdot \boldsymbol{\gamma} ,
\end{align}
which are only functions of $\theta$, not of ${k_{\parallel}}$. In (\ref{tau-2}), the integral over $ \mathrm{k}_{\parallel} $ can be evaluated as $ \int_{0}^{+\infty} \mathrm{k}_{\parallel}^2 e^{-2 k_{\parallel} z_0} d \mathrm{k}_{\parallel} = 1 / 4 z_0^3 $; hence, we obtain
\begin{align}\label{tau-3}
	\hat{ \boldsymbol{\tau}}_r = \frac{ 1 }{ 4z_0^3 \epsilon_0 (2\pi)^2 }  \mathrm{Re} \left \{   \int_{\theta = 0}^{ \theta = 2\pi } d\theta  a_{\theta} \omega_{ \theta }  \left[\boldsymbol{\gamma}^* \times \mathbf{M}_{\theta}\right]  \frac{1} { \omega_{\theta} - \omega_0}  \right \}.
\end{align}

Finally, the integral over $ \theta $ may be written as the corresponding principal value ($\mathcal{P.V.}$) integral plus the contribution of the two poles $\theta = \pm \theta_0$, for which the plasmon frequency in Eq. (\ref{w_theta}) matches the transition frequency of the atom (${\omega _{{\pm \theta_0}}=\omega_0}$),
\begin{align}\label{tau-res}
	\hat{ \boldsymbol{\tau}}_r = \frac{1}{ 4z_0^3 \epsilon_0 (2\pi)^2 }  \mathrm{Re} \left \{ \mathcal{P.V.}   \int_{\theta = 0}^{ \theta = 2\pi }   a_{\theta} \omega_{ \theta }    \frac{ \left[\boldsymbol{\gamma}^* \times \mathbf{M}_{\theta}\right] } { \omega_{\theta} - \omega_0} d\theta  \right. \nonumber \\
	\left.  +   \frac{i\pi   a_{\theta} \omega_{ \theta }} { | \partial_{\theta} \omega_{\theta} | } {\left.
		\right|_{\theta  =
			{\theta _0}}}\  \left[\boldsymbol{\gamma}^* \times \mathbf{M}_{\theta = +\theta_0} + \boldsymbol{\gamma}^* \times \mathbf{M}_{\theta = -\theta_0}  \right] \right \}.
\end{align}


\subsection{Non-resonant term}

The non-resonant (Casimir) component of the torque is given by the second of Eqs. (\ref{tau}). By rewriting it for a single mode, and following the same procedure as for the resonant term, it can be shown that
\begin{align}\label{tau-non}
	\hat{ \boldsymbol{\tau}}_n = \frac{1 }{ 4z_0^3 \epsilon_0 (2\pi)^2 }  \mathrm{Re} \left \{   \int_{\theta = 0}^{ \theta = 2\pi } d\theta  a_{\theta} \omega_{ \theta }  \frac{\boldsymbol{\gamma} \times \mathbf{M'}_{\theta}} { \omega_{\theta} + \omega_0}  \right \},
\end{align}
where the the principal value is not necessary due to the absence of poles on the integration contour, and
\begin{align}
	\mathbf{M'}_{\theta} &= \frac{1}{k_{\parallel}^2} ( i\mathbf{k}_{\parallel} - \mathrm{k}_{\parallel} \hat{z} )( -i\mathbf{k}_{\parallel} - \mathrm{k}_{\parallel} \hat{z} ) \cdot \boldsymbol{\gamma}^*  .
\end{align}


\section{Symmetries, equilibrium states, and stable/unstable phases}
\label{Numerical_results}

\subsection{Optical torque and symmetry considerations}

As discussed in the previous sections (see Eq. (\ref{tau_hat})), the optical torque is composed of two components: (1) a resonant part, $ \hat{\boldsymbol{\tau}}_r $, which is the dominant term when the two-level atom is in its excited energy state, $\left | e \right >  $ (i.e., $  \rho_{ee}(t) = 1 $); and (2) a non-resonant part, $\hat{\boldsymbol{\tau}}_n$, also known as Casimir torque. The non-resonant part becomes dominant once the atom decays to its ground state $  \left | g \right >  $ (i.e., $  \rho_{ee}(t) = 0 $ ). At any intermediate state (i.e., $ 1 > \rho_{ee}(t) > 0 $) both parts contribute. In this section, to illustrate the application and validity of the developed theory, we consider a system composed of a two-level atom with transition frequency $ \omega_0 $ above an interface between vacuum and a magnetized plasma with $ \omega_c / \omega_p = 0.4 $. We compare the exact and quasi-static solutions for the optical torque exerted on the atom, and offer relevant physical insight into some interesting and counterintuitive physical effects.

\begin{figure}[h!]
	\begin{center}
		\noindent \includegraphics[width=3.5in]{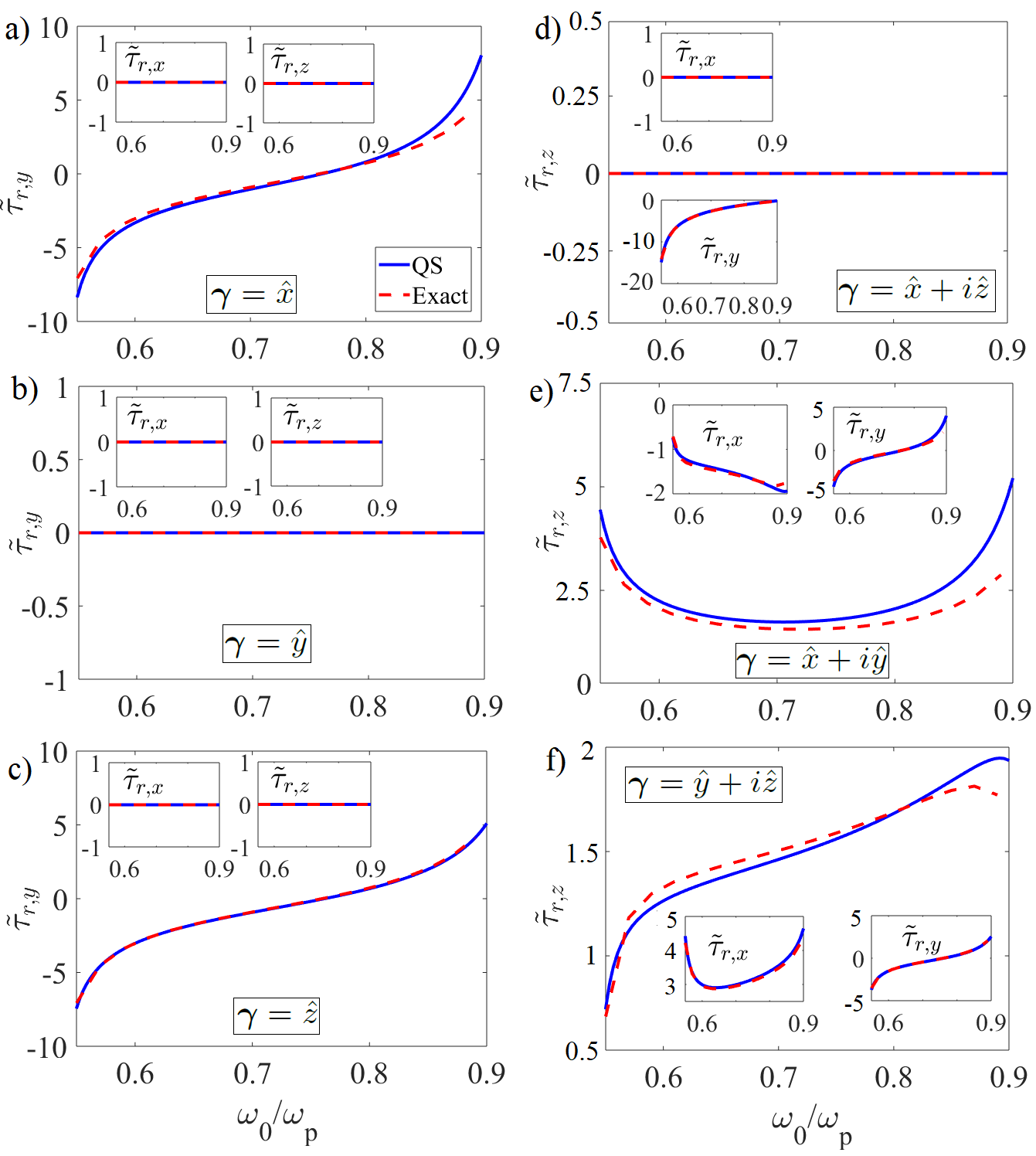}
	\end{center}
	\caption{Resonant part of the optical torque on an initially-excited atom located at $ z_0 = 0.01c/ \omega_p $ in the vacuum region above the plasma-vacuum interface, with $ \omega_c / \omega_p = 0.4 $. The atom radiates at frequency $ \omega_0 $, which is varied within the frequency range $ [ \omega_- ~ \omega_+ ] $, given by Eq. (\ref{w+_w-}). Left column: linearly polarized atoms, with a. $\boldsymbol{\gamma} = \hat{ x}$, b. $\boldsymbol{\gamma} = \hat{ y}$, and c. $\boldsymbol{\gamma} = \hat{ z}$. Right column: circularly polarized atoms, with d. $\boldsymbol{\gamma} = \hat{ x} + i \hat{ z}$, e. $\boldsymbol{\gamma} = \hat{ x} + i \hat{y}$ and f. $\boldsymbol{\gamma} = \hat{ y} + i \hat{z}$.}
	\label{x-l}
\end{figure}

We consider first the resonant component of the optical torque: the exact solution can be obtained from Eq. (\ref{tau_exact}) by assuming the atom to be in its excited state (further details on the calculations of the ``electric part'' of the Green function $\boldsymbol{\mathrm{G}}_{\rm{EE}}$ for a gyrotropic half-space are provided in Appendix \ref{ApGreenHalfSpace}). The quasi-static solution is given by Eq. (\ref{tau-res}). Fig. \ref{x-l} compares the resonant torque, normalized to $ \mathcal{N} = |\gamma|^2/ 16z_0^3 \epsilon_0 \pi $ ($ \tilde{\tau}_j = \hat{ {\tau}}_j / \mathcal{N}$, with $j=x,y,z $), obtained from the exact and quasi-static methods, for different orientations of a linearly- (left column) and circularly- (right column) polarized atom. As clearly seen in the figure, the quasi-static and exact results are in good agreement, which confirms the validity of the assumptions of our quasi-static analysis, namely, the fact that the dynamics of the system is governed by the unidirectional excitation of surface waves on the nonreciprocal material-vacuum interface.

Interestingly, Fig. \ref{x-l} (left column) shows that for a linearly polarized atom the optical torque is non-zero only in certain directions; in particular, the $y$-axis (direction of the bias) seems to be a privileged direction for the torque. These results are consistent with general symmetry considerations applied to the considered system in Fig. \ref{geom} biased by a magnetic field along the $+y$-axis. Due to the presence of a static bias, the rotational symmetry of the system is broken. However, our biased system is symmetric (invariant) under a space inversion (parity transformation) along the $y$-axis, $\mathcal{P}_y:y \rightarrow -y$, which transforms a generic magnetic field (an axial vector) as $(B_x, B_y, B_z) \rightarrow (-B_x, B_y, -B_z)$ (hence, it does not flip the sign of the $y$-directed magnetic bias considered in our system).
%
%
In addition, the operator $\mathcal{P}_y$ transforms the electric dipole moment (a polar vector) as $(\gamma_x, \gamma_y, \gamma_z) \rightarrow (\gamma_x, -\gamma_y, \gamma_z)$. The electric field is transformed in the same way. Hence the torque ($ \propto \boldsymbol{\gamma} \times \textbf{E} $) is transformed as $(\tau_x,~ \tau_y,~ \tau_z) \rightarrow (- \tau_x, ~\tau_y,~ - \tau_z)$. 

Now we consider the particular case of a linear dipole aligned along one of the principal coordinate axes $(x, ~y,~ \mathrm{or} ~z)$, as in the first column of Fig. \ref{x-l}. It is clear that after the transformation described above the atom polarization stays invariant (flipping the sign of the dipole moment does not change the linear polarization state). Due to the system's symmetries, the torque must be invariant under the parity transformation, $\mathcal{P}_y \hat{\boldsymbol{\tau}}  = \hat{\boldsymbol{\tau}} $. Thus, the components of the torque $\tau_x$ and $\tau_z$, which are odd under the transformation, must vanish for a dipole aligned along one of the coordinate axes, meaning that the only non-trivial torque component for a linearly polarized atom is $ \tau_y $. This fact is indeed consistent with the results in Fig. \ref{x-l}. 

\begin{figure}[!htbp]
	\begin{center}
		\noindent \includegraphics[width=3.5in]{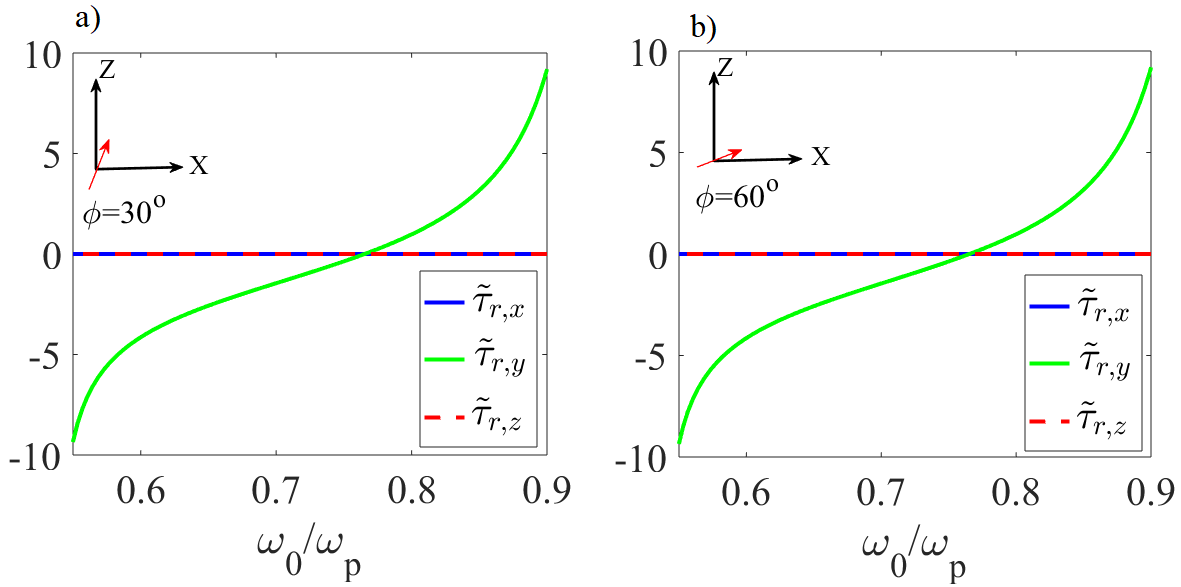}
	\end{center}
	\caption{Optical torque on an initially-excited atom linearly-polarized in the $xoz$ plane, sweeping the radiation frequency in the range $ [ \omega_- ~ \omega_+ ] $ (the other parameters of the system are given in the caption of Fig. \ref{x-l}). The red arrow indicates the orientation of the dipole with respect to the $z$-axis. As discussed in the text, only the $y$ component of the torque is non-zero, keeping the atom in the $xoz$ plane.}
	\label{xoz_L}
\end{figure}

We would like to note that the same symmetry considerations also apply to \emph{any} polarization state completely contained in the $xoz$ plane, predicting that the optical torque is non-zero only along the $y$ direction. This is consistent with the results in Fig. \ref{x-l}-d for the case of a circularly-polarized dipole in the $xoz$ plane. To further confirm these predictions, we have also calculated the torque on an atom linearly polarized in the $xoz$ plane with different angles with respect to the $z$-axis, as shown in Fig. \ref{xoz_L}. Again, the only non-zero component of the torque is $ \tau_y $. All these results indicate that the optical torque will not change the plane of polarization for an atom \emph{arbitrarily} polarized in the $xoz$ plane.

\begin{figure}[h!]
	\begin{center}
		\noindent \includegraphics[width=3.2in]{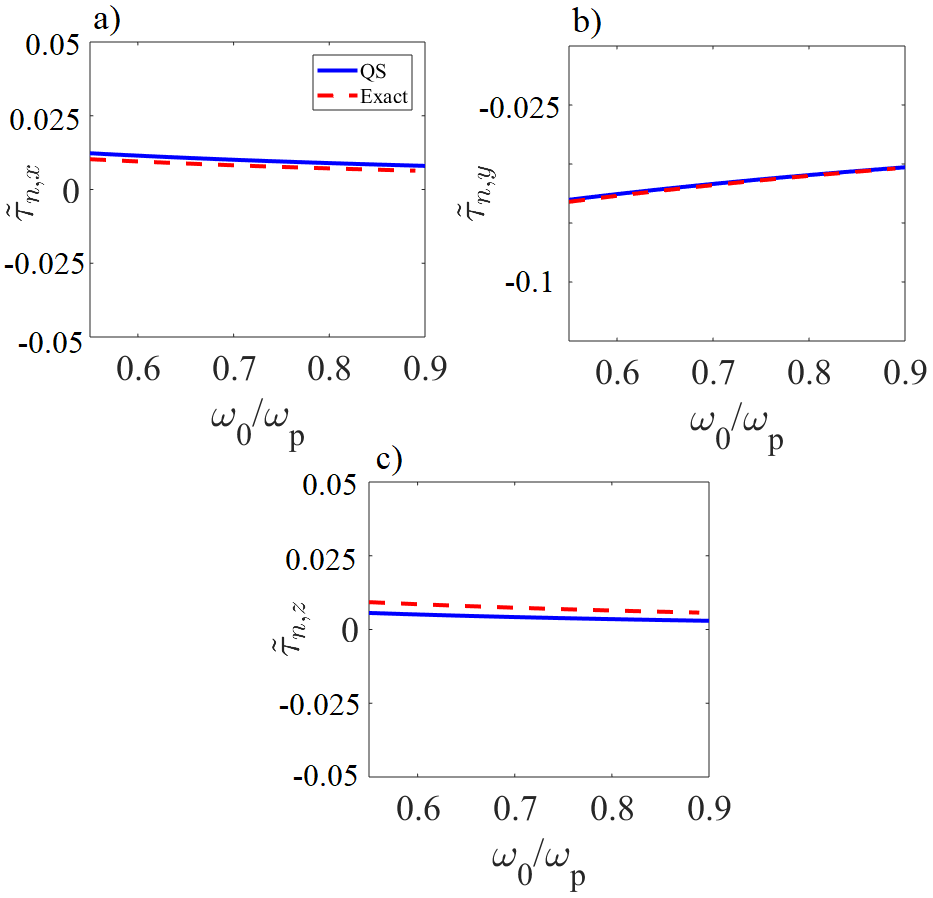}
	\end{center}
	\caption{Non-resonant part of the optical torque (fluctuation-induced Casimir torque) on an atom in its ground state, with $  \boldsymbol{\gamma} = (1 + 1i)\hat{ x} + i \hat{ y} + \hat{ z} $, sweeping the radiation frequency in the range $ [ \omega_- ~ \omega_+ ] $ (the other parameters of the system are given in the caption of Fig. \ref{x-l}). The three panels show the three Cartesian components of the torque, comparing the quasi-static solution (solid blue) with the exact solution (dashed red).}
	\label{infty}
\end{figure}

For the non-resonant component of the optical torque (Casimir torque), the exact solution is given by Eq. (\ref{tau_exact}) specialized for an atom in its ground state (hence, for a sufficiently large time at which the spontaneous emission process is completed). The quasi-static solution is given by Eq. (\ref{tau-non}). We have verified that the exact and quasi-static solutions are in good agreement for all the configurations considered in Fig. \ref{x-l} (linearly-polarized and circularly-polarized atoms along the main axes/planes of the system); for these cases the non-resonant part of the torque is quite small. To further show the general applicability of our theory, we also considered a more complicated (elliptical and oblique) dipole polarization, as indicated in the caption of Fig. \ref{infty}, which yields a larger Casimir torque. As shown in Fig. \ref{infty}, the two solutions are in good agreement, confirming again the validity of our quasi-static assumptions, namely, the fact that response is dominated by unidirectional surface waves. 
%
%
%

\subsection{Polarization equilibrium states for \\ resonant optical torque}

One of the most interesting aspects that can be studied with our theory is the existence of equilibrium states for the polarization of an atom under resonant and non-resonant torque action. In other words, are there planes and axes along which the atom dipole moment tends to spontaneously align?

As discussed in the previous section, the symmetries of the system imply that, for an atom arbitrarily polarized in the $ xoz $ plane, the only non-trivial torque component is along the $y$-axis; therefore, the torque acts to rotate the dipole in the $xoz$ plane, but it does not change the plane of polarization.

\begin{figure}[!htbp]
	\begin{center}
		\noindent \includegraphics[width=3.5in]{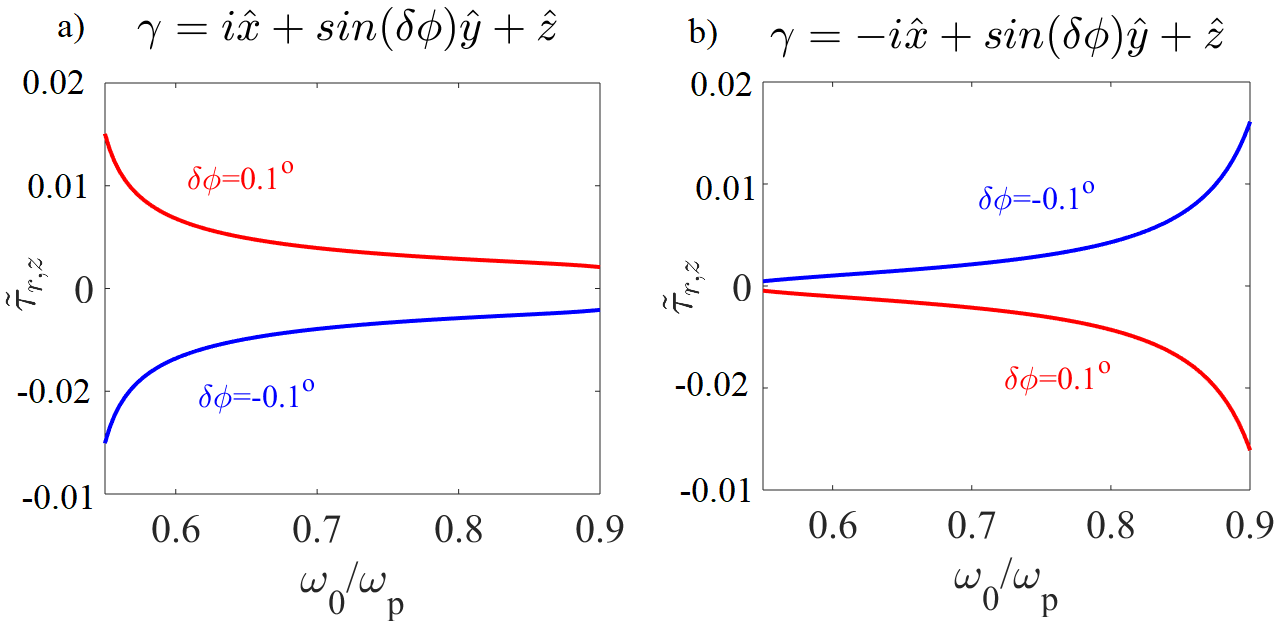}
	\end{center}
	\caption{Normalized optical torque $ \tau_z $, in the frequency range $ [ \omega_- ~ \omega_+ ] $, exerted on an LCP atom (panel a.), and an RCP atom (panel b.), with polarization plane tilted from the $xoz$ plane by a small angle $\delta \phi$. The torque tends to align the polarization plane of an LCP atom to the $xoz$ plane, whereas it tends to rotate the polarization plane of an RCP atom out of the $xoz$ plane.}
	\label{stable}
\end{figure}

To understand whether the $ xoz $ plane truly represents a stable plane for an atom in the environment considered in Fig. \ref{geom}, we perturb the initial polarization state out of this plane. Figure \ref{stable}-a shows a right-handed circularly-polarized (RCP) atom (``right-handed'' looking at the atom toward the +$y$-axis) with dipole moment mostly in the $xoz$ plane, but with a small component along the $y$-axis. When the $y$-component of the dipole moment is positive (negative), a positive (negative) torque appears along the $z$-axis, which tends to rotate the plane of polarization \emph{out} of the $xoz$ plane. Therefore, the $xoz$ plane is not a stable equilibrium plane for an RCP atom. The situation is drastically different for a left-handed circularly-polarized (LCP) atom (``left-handed'' looking at the atom toward the +$y$-axis). As depicted in Fig. \ref{stable}-b, when the LCP atom dipole moment is deviated toward the positive (negative) sides of the $y$-axis, a negative (positive) torque along the $z$-axis appears, which pushes back the plane of polarization to its equilibrium state, hence trying to keep the polarization in the $xoz$ plane. To give an intuitive description of this effect, we note that this behavior is consistent with an heuristic analogy between a CP dipole, a circulating current, and a magnetic dipole: the equivalent magnetic dipole, orthogonal to the plane of polarization of the CP dipole, tends to align along the static magnetic field ($+y$-axis). Depending on the sense of rotation of the CP dipole, the equivalent magnetic dipole will therefore be mostly parallel (LCP) or anti-parallel (RCP) with respect to the bias direction. In the latter case, the torque will make the atom rotate by $ 180^{\mathrm{o}} $ about the $z$-axis, so that the dipole sense of rotation also flips from the point of view of the bias field direction. 

Before proceeding in our discussion, it is worthwhile to briefly summarize the main results obtained so far. If the atom is free to rotate: (i) for any polarization, an atom with dipole moment in the $xoz$ plane will stay in the $xoz$ plane; (ii) a linearly-polarized atom with dipole moment along the bias will remain along the bias; (iii) an arbitrarily-oriented LCP atom will be pushed into the $xoz$ plane. These findings hold over the entire frequency range where unidirectional SPPs exist, i.e., $  \omega_- \leq \omega \leq \omega_+ $

\begin{figure}[!htbp]
	\begin{center}
		\noindent \includegraphics[width=3.5in]{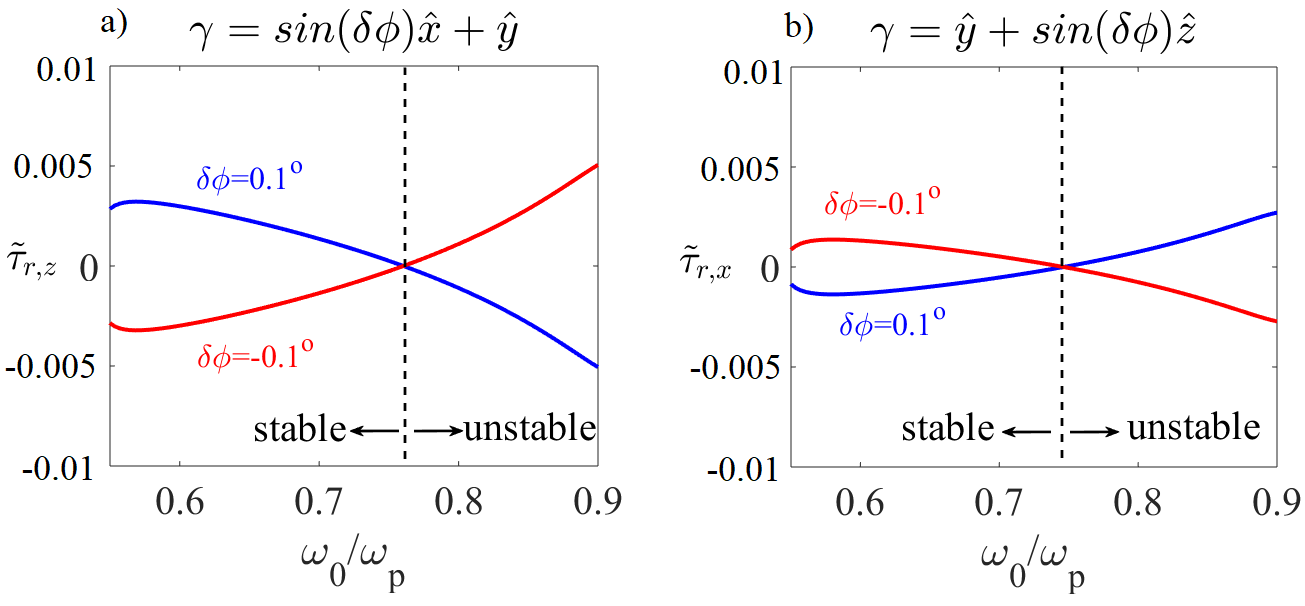}
	\end{center}
	\caption{Normalized optical torque components, $ \tau_z $ (panel a.) and $ \tau_x $ (panel b.), in the frequency range $ [ \omega_- ~ \omega_+ ] $, exerted on an atom linearly polarized in a direction slightly deviated from the $y$-axis by a small angle $\delta \phi$ toward a. the $x$-axis, and b. the $z$-axis. Depending on the frequency, an atom with linear polarization nearly parallel to the $y$-axis (direction of the static bias) may exhibit stable or unstable behavior. The vertical dashed line represents the frequency boundary between stable and unstable phases, corresponding to the vertical dashed line in Fig. (\ref{w_t}).}
	\label{stable_linear}
\end{figure}

Next, we study whether the direction of the bias itself provides a stable equilibrium position for a linearly-polarized atom in this frequency range. As shown in Fig. \ref{x-l}-b, when the atom is linearly-polarized along the $y$-axis, the torque is identically zero. To investigate the stability of an atom along this direction, we deviate the polarization along the $x$- and $z$-axis, and study the resulting torque on the atom. Figure \ref{stable_linear}-a shows the case in which the polarization has a small positive (negative) angle with respect to the $+x$-axis. We see that, for the range of frequencies 
\begin{equation} \label{stable_range}
	\omega_- = \frac{1}{2} \left( -\omega_c + \sqrt{ 2\omega_p^2 + \omega_c^2 }  \right) < \omega_0 < \omega_m = \sqrt{ \frac{\omega_p^2}{2}  + \frac{\omega_c^2}{2} },
\end{equation}
a positive (negative) torque along the $z$-axis appears, which pushes the polarization back toward the $y$-axis; conversely, for the rest of the frequency range, the sign of the torque flips, which tends to deviate the atom polarization even more toward the $x$-axis. Similarly, if we consider a small deviation along the $z$-axis, a non-zero torque appears along the $x$-axis, as shown in Fig. \ref{stable_linear}-b. In this case, within the same frequency range as in (\ref{stable_range}), a positive (negative) torque appears for negative (positive) deviations, which brings back the atom to its initial state along the $y$-axis, whereas for the rest of the frequency range the atom polarization is unstable and the torque tries to increase the $z$-component of the dipole moment. Therefore, for a linearly-polarized atom along the $y$-axis (direction of magnetic bias), for transition frequencies within $ [\omega_-~ \omega_m] $,  \emph{we observe an unusually stable behavior: the linearly-polarized electric dipole tends to align itself along the static magnetic field direction, which is a behavior typically expected from magnetic dipoles}. Conversely, for transition frequencies larger than $ \omega_m $ the state of polarization along the $y$-axis is unstable. The boundary between these two regimes is the frequency $\omega_m$, at which the momentum of the SPP modes becomes parallel to the $y$-axis (bias direction) with $\theta= \pm 90^{\mathrm{o}}$, as seen in Eq. (\ref{w_theta}) and Fig. \ref{w_t} (dashed black line in Fig. \ref{w_t}). Exactly at this frequency the atom releases momentum symmetrically with respect to the geometry of the system, hence the recoil force felt by the atom \cite{PRA_force,PRL_force} is minimized. Interestingly, by tuning the bias, and thereby the cyclotron frequency, this boundary frequency $\omega_m$ between stable and unstable phases can be largely tuned, which provides an additional degree of freedom for the manipulation of small polarized objects.

\subsection{Polarization equilibrium states for fluctuation-induced Casimir torque}

\begin{figure}[!htbp]
	\begin{center}
		\noindent \includegraphics[width=3.4in]{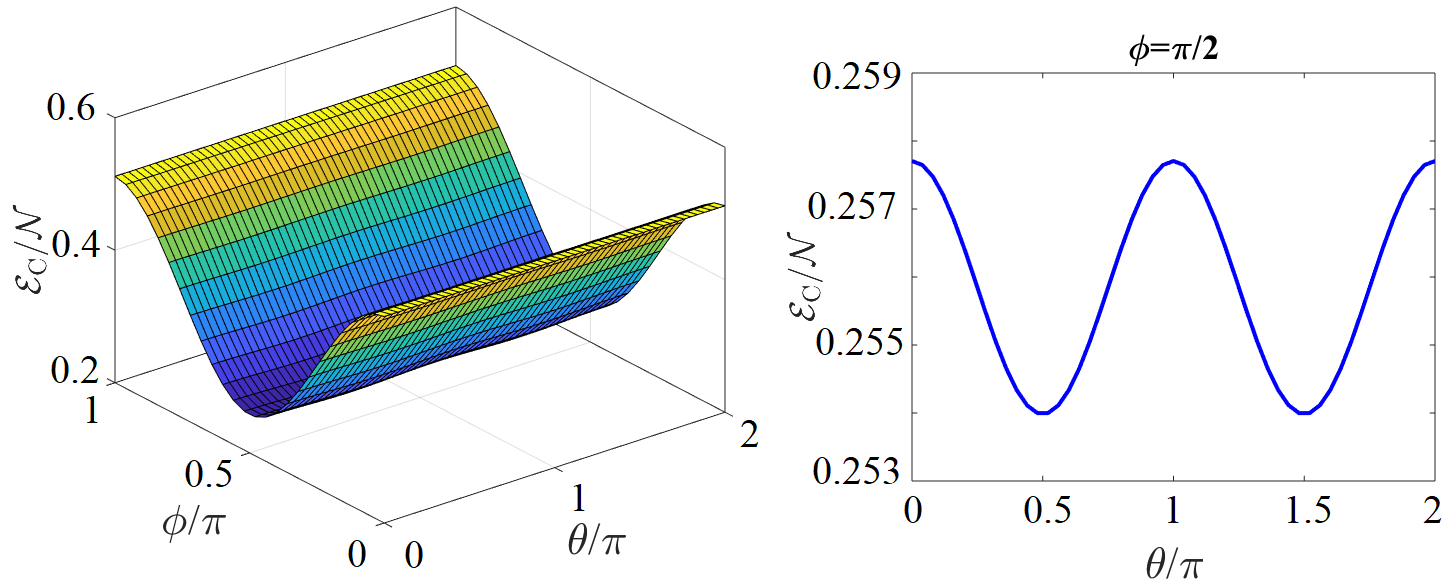}
	\end{center}
	\caption{Normalized zero-point interaction energy for a linearly polarized atom above a nonreciprocal (gyrotropic) material with properties given in the caption of Fig. \ref{x-l}, for different angular orientations. As indicated in Fig. \ref{geom}, $ \theta $ is measured from the $+x$-axis and $ \phi $ from the $ +z $-axis. The derivative of the zero-point energy, with respect to the angle, gives the Casimir torque on the atom, as shown in Appendix \ref{zero_point_lossy}. }
	\label{EC}
\end{figure}

As shown in Appendix \ref{zero_point_lossy}, the Casimir torque (non-resonant part of the torque) can be calculated directly from the zero-point interaction energy \cite{PRA_force}
\begin{align} \label{zero_point_lossy_eq}
	\mathcal{E}_C =  - \frac{1}{{2\pi }}\int\limits_0^\infty  {d\xi {\rm{ }}} \,\, {\rm{tr}}\left\{ {{{\left( { - i\omega \overline {\bf{G}} } \right)}_{\omega  = i\xi }} \cdot \tilde \alpha \left( {i\xi } \right)} \right\}
	\end{align}
by taking the derivative of the energy with respect to the spatial angle of the dipole (the energy needed to rotate an object by an angle $\varphi$  is $\tau \varphi$). Eq. (\ref{zero_point_lossy_eq}) is valid for both lossless and lossy environments, and the quantity  $\tilde \alpha$ is the normalized polarizability of the two-level system. Further details are discussed in Appendix \ref{zero_point_lossy}. Hence, since the Casimir torque is directly determined by the angular distribution of the zero-point energy, it is possible to find the equilibrium states of the polarization for Casimir torque by plotting the energy as a function of the orientation of the dipole. This is done in Fig. \ref{EC}, where the minima of the energy corresponds to the stable equilibrium positions and the maxima of the energy corresponds to unstable equilibrium points. These results show that an atom \emph{in its ground state} above a nonreciprocal gyrotropic material will experience a non-zero torque if its orientation is not along one of the stable directions in Fig. \ref{EC}, namely, $ \phi = \pi / 2,~ \theta = \pi/2 ~ \mathrm{or} ~ 3\pi/2 $, corresponding to the $y$-axis of the system (bias direction). Hence, if the atom is free to move, it will gain a finite non-zero amount of energy, in the form of kinetic energy, corresponding to the difference between its initial state and the final state, i.e., the minima of the Casimir energy in Fig. \ref{EC}. This energy could then be released, exactly one time, for example in the form of a photon or phonon, and the atom would align itself along the $y$-axis to minimize the zero-point energy (note that this ``extraction'' of energy from the quantum vacuum does not violate any thermodynamics law: this finite amount of energy always comes from the work done to prepare the system in a configuration where Casimir force/torque is observed).


\section{Conclusion}

\label{SectConcl}

In this paper, we have investigated the quantum optical torque acting on a two-level system, initially prepared in an arbitrary state, interacting with a general (nonreciprocal, bianisotropic, and dispersive) electromagnetic environment under the Markov approximation. We have rigorously shown that the optical torque can be decomposed into resonant and non-resonant parts, expressed explicitly in terms of the system's Green function. When the atom is initially excited, and undergoes a spontaneous-emission process, the resonant term dominates, governed by the resonantly-excited electromagnetic modes of the surrounding structure (e.g., guided surface modes if the atom is above a stratified medium);  conversely, when the atom decays to its ground state, it is the non-resonant part of the torque (fluctuation-induced Casimir torque) that dominates the response.

We have then applied our general theory to study the relevant case of a two-level atom above a continuous material with strong nonreciprocity, implemented in the form of an electric plasma biased by a static magnetic field. For this physical system, the optical torque has been evaluated with our exact formulation and with an approximated quasi-static analysis. Numerical studies confirm the emergence of non-zero torque on the atom in specific directions, due to the excitation of unidirectional surface plasmons on the nonreciprocal material interface. Interestingly, we have discovered the existence of equilibrium polarization planes and axes, determined by the direction of the magnetic bias, along which the polarized atom naturally tends to align, behaving analogously to a magnetic dipole in a constant magnetic field. We have also found that, depending on the atom energy state, polarization, and atomic transition frequency, there are distinct stable and unstable phases, in which the optical torque produces a force working to change the polarization of the atom (unstable) or to keep the atom in its initial polarization state (stable). 

We expect that the theoretical predictions presented in this paper would still be valid, at least qualitatively, in other types of structures that support unidirectional wave propagation, as the described physical effects mostly depend on the existence of a preferential sense of direction in the system, rather than on its specific implementation details. In this context, we would like to note that our general theory can be applied to study other classes of nonreciprocal and topological systems, and, more broadly, any inhomogeneous, dissipative, dispersive, bianisotropic structure.

In summary, we believe that the theory and results presented in this paper offer new relevant physical insight into the electrodynamics of quantum emitters near complex media, and may open new interesting research directions at the intersection of different fields, including nanophotonics, quantum optics, opto-mechanics, and the emerging area of topological and nonreciprocal photonics.



\section*{Acknowledgments}

S.A.H.G. and F.M. acknowledge support from the National Science Foundation (NSF) with Grant No. 1741694. M. S. acknowledges support from Funda\c{c}\~{a}o para Ci\^{e}ncia e a Tecnologia (FCT) under projects PTDC/EEITEL/ 4543/2014 and UID/EEA/50008/2013.

\begin{widetext}
	
	\appendix
	
	\section{Electric dyadic Green function for a three-dimensional nonreciprocal material half-space}
	
	\label{ApGreenHalfSpace}
	The electric dyadic Green function $\mathbf{ G}_{\mathrm{EE}}$ relates the electric field emitted by a classical dipole with its electric dipole moment $\boldsymbol{\gamma}$ through the relation $ \mathbf{ E} = \mathbf{ E}^p +\mathbf{ E}^s = -i\omega \mathbf{ G}_{\mathrm{EE}}  \cdot \boldsymbol{\gamma}  $, where $\mathbf{ E}^p$ is the primary field radiated by the dipole in free space, and $\mathbf{ E}^s$ is the scattered field from the environment in which the dipole is located. Since the expression of $\mathbf{ E}^p$ is given in several textbooks, we only derive here the scattered field for the structure of interest. For an arbitrary-polarized electric dipole (emitting atom) above the interface between a gyrotropic medium (magnetized plasma) and vacuum, the scattered electric field above the interface can be written as \cite{PRA_force} 
	\begin{equation}\label{Es_exact}
	\mathbf{E}^{s}=\frac{1}{(2\pi )^{2}}\int \int d {k}_{x}d {k}%
	_{y}e^{i\mathbf{k}_{\Vert }\cdot \mathbf{r}}\frac{e^{-\gamma _{0}(d+z)}}{%
		2\gamma _{0}}\mathbf{C}\left( \omega ,\mathbf{k}_{\Vert }\right) \cdot \frac{%
		\mathbf{\gamma }}{\varepsilon _{0}},
	\end{equation}%
	where
	\begin{align}\label{C}
	& \mathbf{C}\left( \omega ,\mathbf{k}_{\Vert }\right) =\left( \mathbf{I}_{\Vert }+\widehat{\mathbf{z}}\frac{i\mathbf{k}_{\Vert }%
	}{\gamma _{0}}\right) \cdot \boldsymbol{\mathrm{R}}\left( \omega ,\mathbf{k}%
	_{\Vert }\right) \cdot \left( i\gamma _{0}\mathbf{k}_{\Vert }\widehat{%
		\mathbf{z}}+k_{0}^{2}\mathbf{I}_{\Vert }-\mathbf{k}_{\Vert }\mathbf{k}%
	_{\Vert }\right) ,
	\end{align}%
	with $\mathbf{I}_{\Vert }=\widehat{\mathbf{x}}\widehat{\mathbf{x}}+\widehat{%
		\mathbf{y}}\widehat{\mathbf{y}}$ and $\boldsymbol{\mathrm{k}}_{\parallel }=%
	{k}_{x}\mathbf{\hat{x}}+{k}_{y}\mathbf{\hat{y}}$. Here, ${\mathbf{%
			R}(\omega ,{k_{x}},{k_{y}})}$ is a $2\times 2$ reflection matrix that relates the tangential (to the interface) components $x$ and $y$ of the reflected electric field to the corresponding $x$ and $y$
	components of the incident field \cite{PRA_force}
	\begin{equation} 
	\boldsymbol{\mathrm{R}} \left( \omega ,\mathbf{k}_{\Vert }\right) = \left(\boldsymbol{\mathrm{Y}}_0 + \boldsymbol{%
		\mathrm{Y}}_g\right)^{-1} \cdot \left( \boldsymbol{\mathrm{Y}}_0 -
	\boldsymbol{\mathrm{Y}}_g \right),
	\end{equation}
	where 
	\begin{align}
	& \boldsymbol{\mathrm{Y}}_0 = \frac{1}{ i {k}_0 \gamma_0 } \left(
	\begin{array}{cc}
	-\gamma_0^2 + {k}_x^2 & {k}_x {k}_y \\
	{k}_x {k}_y & -\gamma_0^2 + {k}_y^2%
	\end{array}%
	\right),
	\end{align}
	and 
	\begin{align}
	&\boldsymbol{\mathrm{Y}}_g = \left(%
	\begin{array}{cc}
	\frac{\Delta_1 {k}_{t,1}^2 } {{k}_0} & \frac{\Delta_2 {k%
		}_{t,2}^2 } {{k}_0} \\
	\frac{\Delta_1 {k}_x {k}_y + i \gamma_{z,1} ( \theta_1 -1 ) {k}_y}{
		{k}_0 } & \frac{\Delta_2 {k}_x {k}_y + i
		\gamma_{z,2} ( \theta_2 -1 ) {k}_y}{ {k}_0 }%
	\end{array}%
	\right)   \cdot \left(%
	\begin{array}{cc}
	{k}_x + i \gamma_{z,1} \Delta_1 & {k}_x + i \gamma_{z,2}
	\Delta_2 \\
	\theta_1 {k}_y & \theta_2 {k}_y%
	\end{array}%
	\right)^{-1}
	\end{align}
	relates the tangential components of the magnetic field to the tangential components of the electric field in vacuum and in the magnetized plasma, respectively. The parameters $ \gamma_{z,i}, ~ i=1,2 $, with $ \mathrm{Re}(\gamma_{z,i}) > 0 $ and $ \Delta_{i},~ \theta_{i} $, in the matrices above are defined in \cite{PRA_force}.
	
	Equation (\ref{C}) can be written in the following form 
	
	\begin{equation}
	\mathbf{C}\left( \omega ,\mathbf{k}_{\Vert }\right) = \mathbf{A} \cdot \mathbf{R} \left( \omega ,\mathbf{k}_{\Vert }\right) \cdot \mathbf{B},
	\end{equation}
	with 
	
	\begin{equation}
	\mathbf{A} = \hat{x}\hat{x} + \hat{y} \hat{y} + \frac{i}{\gamma_0} ( k_x \hat{z}\hat{x} + k_y \hat{z} \hat{y} ) = \left(\begin{array}{ccc}
	1 & 0 & 0 \\
	0 & 1 & 0 \\
	\frac{ik_x}{\gamma_0} & \frac{ik_y}{\gamma_0} & 0
	\end{array}\right)
	\end{equation}
	
	\begin{align}
	& \mathbf{B} = i\gamma_0 ( k_x \hat{x} \hat{z} + k_y \hat{y} \hat{z}) + k_0^2 \hat{x} \hat{x} + k_0^2 \hat{y} \hat{y} -k_x^2 \hat{x} \hat{x} - k_xk_y \hat{x} \hat{y} - k_yk_x \hat{y} \hat{x} - k_yk_y \hat{y} \hat{y} =  \left(\begin{array}{ccc}
	k_0^2 -k_x^2 & -k_xk_y & i\gamma_0 k_x \\
	-k_yk_x & k_0^2 - k_y^2 & i\gamma_0 k_y \\
	0 & 0 & 0
	\end{array}\right).
	\end{align} 
	
	The reflection matrix $ \mathbf{R} \left( \omega ,\mathbf{k}_{\Vert }\right)  $ is a $ 2 \times 2 $ matrix, which we write in $3 \times 3$ form as 
	
	\begin{equation}
	\mathbf{R} \left( \omega ,\mathbf{k}_{\Vert }\right)  = \left(\begin{array}{ccc}
	R_{11} & R_{12} & 0 \\
	R_{21} & R_{22} & 0\\
	0 & 0 & 0
	\end{array}\right),    
	\end{equation}
	therefore,
	
	\begin{align}
	\mathbf{C} =& \left(\begin{array}{ccc}
	1 & 0 & 0 \\
	0 & 1 & 0 \\
	\frac{ik_x}{\gamma_0} & \frac{ik_y}{\gamma_0} & 0
	\end{array}\right) \cdot  \left(\begin{array}{ccc}
	R_{11} & R_{12} & 0 \\
	R_{21} & R_{22} & 0\\
	0 & 0 & 0
	\end{array}\right)    \cdot  \left(\begin{array}{ccc}
	k_0^2 -k_x^2 & -k_xk_y & i\gamma_0 k_x \\
	-k_yk_x & k_0^2 - k_y^2 & i\gamma_0 k_y \\
	0 & 0 & 0
	\end{array}\right) = \notag \\& 
	 = \left(\begin{array}{ccc}
	R_{11} ( k_0^2 - k_x^2 ) + R_{12} ( -k_yk_x ) & R_{11} (-k_xk_y) + R_{12}( k_0^2-k_y^2 ) & R_{11}(i\gamma_0 k_x) + R_{12} ( i\gamma_0 k_y )  \\ 
	R_{21} ( k_0^2 - k_x^2 ) + R_{22} ( -k_yk_x ) & R_{21} (-k_xk_y) + R_{22}( k_0^2-k_y^2 ) & R_{21}(i\gamma_0 k_x) + R_{22} ( i\gamma_0 k_y ) \\ 
	J_{31} ( k_0^2 - k_x^2 ) + J_{32} ( -k_yk_x ) & J_{31} (-k_xk_y) + J_{32}( k_0^2-k_y^2 ) & J_{31} (i\gamma_0 k_x) + J_{32} ( i\gamma_0 k_y )
	\end{array}\right),
	\end{align}
	where  
	
	\begin{align}
	& J_{31} = \frac{i}{\gamma_0}(  k_xR_{11} + k_y R_{21} ) \notag \\&
	J_{32} = \frac{i}{\gamma_0}(  k_xR_{12} + k_y R_{22} ).
	\end{align}
	
	For an atom with generic polarization, $ \boldsymbol{\gamma} = \gamma_x \hat{x} + \gamma_y \hat{y} + \gamma_z \hat{z} $, the integrand of (\ref{Es_exact}) can therefore be obtained in a straightforward manner from the equations above.
	


\section{Zero-point torque for lossy environment}
\label{zero_point_lossy}

While the Green function for a closed lossless environment is analytic everywhere in the complex frequency plane except on the real axis, in the lossy case the Green function is usually not analytic in the lower half plane, and may have poles on the lower-half-plane imaginary axis. Hence, our integral formulation in the main text only strictly applies to lossless environments. Nevertheless, it is known from the fluctuation-dissipation theorem that fluctuation-induced forces/torques are determined, in all cases, by the values of the Green function in the upper half plane \cite{Lifshitz}. Therefore, if we can write the response (e.g., Casimir torque) of the system in terms of only the upper-half-plane values of the Green function, this modified formulation would apply to both lossless and lossy cases. 

From Eq. (\ref{tau_exact}), the torque for the quantum-vacuum case is 

	\begin{align}\label{tauCasimir}
	{ \boldsymbol{\tau}}_C  =  \,{\mathop{\rm Re}\nolimits} \left\{
	\frac{1}{{2\pi }}\int\limits_{ - \infty }^\infty  {d\xi }
	\left[{{\bf{\tilde \gamma }}^*} \times \frac{{{{\left( { - i\omega
						{{ {\bf{G}} } }} \right)}_{\omega  = i\xi }}}}{{{\omega _0} - i\xi
	}} \cdot {\bf{\tilde \gamma }}  + {{\bf{\tilde \gamma }}} \times
	\frac{{{{\left( { - i\omega {{ {\bf{G}} } }} \right)}_{\omega  =
					i\xi }}}}{{{\omega _0} + i\xi }} \cdot {\bf{\tilde \gamma^*
	}}\right] \right\}.
	\end{align}

This result was derived under the assumption that the system is
lossless. To extend it to a lossy environment, as mentioned above, we need to rewrite
the torque as an integral in the upper-half complex frequency plane.

To this end, still assuming for now that there is no dissipation, we
rewrite the zero-point torque in a more compact manner as
\begin{align}\label{tauC2}
{ \boldsymbol{\tau}}_C  = \frac{1}{{2\pi }}{\mathop{\rm
		Re}\nolimits} \int\limits_{ - \infty }^\infty  {d\xi {\rm{ }}} \,\,
{\rm{tr}}\left( {\tilde \alpha \left( {i\xi } \right) \times
	{{\left( { - i\omega {\bf{G}} } \right)}_{\omega  = i\xi }}} \right),
\end{align}
where
\begin{align}\label{alftil}
\tilde \alpha \left( {i\xi } \right) = \left( {{\bf{\tilde \gamma
	}}{{{\bf{\tilde \gamma }}}^*}\frac{1}{{{\omega _0} - i\xi }} +
	{{{\bf{\tilde \gamma }}}^*}{\bf{\tilde \gamma }}\frac{1}{{{\omega
				_0} + i\xi }}} \right)
\end{align}
is the normalized polarizability of the two-level atom. For two
generic ($6\times6$) matrices $\bf{A}$ and $\bf{B}$, we define
\begin{align}
{\rm{tr}}\left( {{\bf{A}} \times {\bf{B}}} \right) \equiv
\sum\limits_i {{{{\bf{\hat u}}}_i} \cdot \left( {{\bf{A}} \times
		{\bf{B}}} \right) \cdot {{{\bf{\hat u}}}_i}},
\end{align}
where ${\bf{A}} \times {\bf{B}} = {\bf{A}} \cdot \left(
{\begin{array}{*{20}{c}}
	{{\bf{1}} \times }&0\\
	0&{{\bf{1}} \times }
	\end{array}} \right){\bf{B}}$ is a rank 3 tensor, and ${\bf{1}}$ is the unit matrix of dimension 3. Note that ${\rm{tr}}\left( {{\bf{A}} \times {\bf{B}}} \right)$ is
a vector.

It can be checked that $\tilde \alpha \left( {i\xi } \right) =
{\tilde \alpha ^*}\left( {i\xi } \right)$ and, because of the
reality of the electromagnetic field,
\begin{equation} \label{Apreality}
{{\bf{G}}^*}\left( {{\bf{r}},{\bf{r'}},\omega } \right) =
{\bf{G}}\left( {{\bf{r}},{\bf{r'}}, - {\omega ^*}} \right).
\end{equation}
Hence, ${{\bf{G}}}\left(
{{\bf{r}},{\bf{r}},\omega } \right)$ is always real-valued on the
imaginary-frequency axis. These properties show that it is possible
to drop the ``$\rm{Re}$'' operator in Eq. (\ref{tauC2}).

Furthermore, we note that for a lossless system the Green function satisfies
\begin{equation}
{{\bf{G}}^\dag }\left( {{\bf{r}},{\bf{r'}},\omega } \right) =  -
{\bf{G}}\left( {{\bf{r'}},{\bf{r}},{\omega ^*}} \right),
\end{equation}
which, in conjuction with Eq. (\ref{Apreality}), implies that
\begin{equation}
{\left[ { - i\omega {\bf{G}}\left( {{\bf{r}},{\bf{r}},\omega }
		\right)} \right]_{\omega  =  - i\xi }} = {\left[ { - i\omega
		{\bf{G}}\left( {{\bf{r}},{\bf{r}},\omega } \right)}
	\right]^T}_{\omega  =  + i\xi }.
\end{equation}
Using the above formula and $\tilde \alpha \left( { - i\xi } \right)
= {\tilde \alpha ^T}\left( {i\xi } \right)$ in Eq. (\ref{tauC2}), we
can write the zero-point torque as an integral over the semi-infinite section of
the imaginary axis contained in the upper-half plane,

	\begin{equation}
	{ \boldsymbol{\tau}}_C = \frac{1}{{2\pi }}\int\limits_0^\infty {d\xi
		{\rm{ }}} \left[ {{\rm{tr}}\left\{ {\tilde \alpha \left( {i\xi }
			\right) \times {{\left( { - i\omega {\bf{G}}} \right)}_{\omega  =
					i\xi }}} \right\} + {\rm{tr}}\left\{ {{{\tilde \alpha }^T}\left(
			{i\xi } \right) \times {{\left( { - i\omega {\bf{G}}}
					\right)}^T}_{\omega  = i\xi }} \right\}} \right].
	\end{equation}
	Taking into account that ${\rm{tr}}\left( {{\bf{A}} \times
		{\bf{B}}} \right) = - {\rm{tr}}\left( {{{\bf{B}}^T} \times
		{{\bf{A}}^T}} \right)$, we get
	\begin{equation}
	{ \boldsymbol{\tau}}_C = \frac{1}{{2\pi }}\int\limits_0^\infty
	{d\xi {\rm{ }}} \left[ {{\rm{tr}}\left\{ {\tilde \alpha \left( {i\xi
			} \right) \times {{\left( { - i\omega {\bf{G}}} \right)}_{\omega  =
					i\xi }}} \right\} - {\rm{tr}}\left\{ {{{\left( { - i\omega {\bf{G}}}
					\right)}_{\omega  = i\xi }} \times \tilde \alpha \left( {i\xi }
			\right)} \right\}} \right]
	\end{equation}
	Finally, this result may be spelled out as follows
	\begin{align} \label{tauC_final}
	{ \boldsymbol{\tau}}_C & = \frac{1}{{2\pi }}\int\limits_0^\infty
	{d\xi {\rm{ }}} \left( {\frac{1}{{{\omega _0} - i\xi }}{{\bf{\tilde
					\gamma }}^*} \times {{\left( { - i\omega {\bf{G}} } \right)}_{\omega
				= i\xi }} \cdot {\bf{\tilde \gamma }} + \frac{1}{{{\omega _0} + i\xi
		}}{\bf{\tilde \gamma }} \times {{\left( { - i\omega  {\bf{G}} }
				\right)}_{\omega = i\xi }} \cdot {{\bf{\tilde \gamma }}^*}} \right)
	\notag \\ & {\rm{  }} - \frac{1}{{2\pi }}\int\limits_0^\infty  {d\xi
		{\rm{ }}} \left( {\frac{1}{{{\omega _0} - i\xi }}{{\bf{\tilde \gamma
			}}^*} \cdot {{\left( { - i\omega {\bf{G}} } \right)}_{\omega  = i\xi
		}} \times {\bf{\tilde \gamma }} + \frac{1}{{{\omega _0} + i\xi
		}}{\bf{\tilde \gamma }} \cdot {{\left( { - i\omega  {\bf{G}} }
				\right)}_{\omega  = i\xi }} \times {{\bf{\tilde \gamma }}^*}}
	\right).
	\end{align}
	This formula gives the zero-point torque as an integral of the
	system's Green function evaluated in the upper half of the complex frequency plane, and thereby it can be used to evaluate the torque of generic lossy photonic
	systems.
	
	For completeness, next we prove that the zero-point torque [Eq.
	(\ref{tauC_final})] can be directly written in terms of the
	zero-point energy of the system \cite{PRA_force}. The formula given
	in Ref. \cite{PRA_force} is only valid for lossless systems, but it can be
	extended to lossy systems using an approach similar to the one used above
	for the torque, obtaining
	\begin{align}
	{{\mathcal{E}}_{C}} =- \frac{1}{{2\pi }}\int\limits_{ 0}^\infty
	{d\xi {\rm{ }}} \left( {\frac{1}{{{\omega _0} - i\xi
		}}{{{\bf{\tilde \gamma }}}^*} \cdot {{\left( { - i\omega
					{\bf{G}} } \right)}_{\omega  = i\xi }} \cdot {\bf{\tilde
				\gamma }} + \frac{1}{{{\omega _0} + i\xi }}{\bf{\tilde \gamma }} \cdot
		{{\left( { - i\omega  {\bf{G}} } \right)}_{\omega  = i\xi }} \cdot
		{{{\bf{\tilde \gamma }}}^*}} \right),
	\end{align}
	which corresponds to Eq. (\ref{zero_point_lossy_eq}).
	
	As an example, we consider that the atom polarization is linear, and we
	calculate the torque along the $z$-direction. Hence, we can write $
	\boldsymbol{\gamma} = \gamma ( \mathrm{cos}(\theta),~
	\mathrm{sin}(\theta), ~ 0 ) $ ($ \theta $ is measured from the
	$+x$-axis, see Fig. \ref{geom}). Since $ \partial_{\theta} = \gamma
	( -\mathrm{sin}(\theta),~ \mathrm{cos}(\theta), ~ 0 ) = \hat{
		\boldsymbol{z}} \times \boldsymbol{\gamma}  $, and noting that the
	Green function is independent of the dipole orientation, it follows
	that
	\begin{align}
	-\partial_{\theta} \mathcal{E}_C = &   \frac{1}{{2\pi
	}}\int\limits_{ 0 }^\infty  {d\xi {\rm{ }}} \left(
	{\frac{1}{{{\omega _0} - i\xi
		}} \hat{ \boldsymbol{z}}\times {{{\bf{\tilde \gamma }}}^*} \cdot {{\left( { - i\omega
					{\bf{G}} } \right)}_{\omega  = i\xi }} \cdot {\bf{\tilde
				\gamma }} + \frac{1}{{{\omega _0} + i\xi }}  \hat{ \boldsymbol{z}}\times {\bf{\tilde \gamma }} \cdot
		{{\left( { - i\omega  {\bf{G}} } \right)}_{\omega  = i\xi }} \cdot
		{{{\bf{\tilde \gamma }}}^*}} \right) \notag \\ &
	+ \frac{1}{{2\pi }}\int\limits_{ 0
	}^\infty  {d\xi {\rm{ }}} \left( {\frac{1}{{{\omega _0} - i\xi
		}}  {{{\bf{\tilde \gamma }}}^*} \cdot {{\left( { - i\omega
					{\bf{G}} } \right)}_{\omega  = i\xi }} \cdot \hat{ \boldsymbol{z}}\times  {\bf{\tilde
				\gamma }} + \frac{1}{{{\omega _0} + i\xi }}  {\bf{\tilde \gamma }} \cdot
		{{\left( { - i\omega  {\bf{G}} } \right)}_{\omega  = i\xi }} \cdot
		{\hat{ \boldsymbol{z}}\times {{\bf{\tilde \gamma }}}^*}}
	\right).
	\end{align}
	By comparing this formula with Eq. (\ref{tauC_final}), we see that,
	as expected, $\hat{ \boldsymbol{z}} \cdot {\boldsymbol{\tau}}_C =
	-\partial_{\theta} \mathcal{E}_C$.

\end{widetext}


\end{document}